\documentclass[12pt]{article}
\usepackage{epsfig}
\begin{document}
\begin{center}
{\Large \textbf{Parameters of the Disk Loaded Waveguide structure for intermediate particles acceleration 
in the intermediate energy range.}}\\
\vspace*{0.8cm}
V.V. Paramonov \\
Institute for Nuclear Research, 117312, Moscow, Russia
\end{center}
\begin{abstract} 
      The Disk Loaded Waveguide (DLW) is the mostly used high frequency structure for 
acceleration of lightweight particles - electrons in the high energy range. In some physical 
experiments acceleration of more heavy particles - muons to medium energies $\gamma \sim 3$ is 
required. DLW parameters are considered for  particle velocity $0.04 \leq \beta \leq 1$ both 
for the fundamental and the nearest backward spatial harmonics. Physical and technical 
restrictions for DLW application in the low $\beta$ range and lower frequency (the L-band range) 
 are analyzed. Basing on particularities of acceleration with Traveling Wave (TW) mode, deep 
optimization of DLW cells dimensions, the choice of optimal operating phase advance for 
each DLW section and combination of forward and backward TW modes, it is possible to create 
simple, cost effective acceleration system for acceleration in the velocity range 
$0.2 \leq \beta \leq 1$ for intermediate particles, in some parameters overcoming accelerating system 
with RF cavities in Standing Wave (SW) mode. Design criteria are 
discussed and examples of possible accelerating systems are presented.
\end{abstract}
\newpage
\tableofcontents
\newpage
\section{Introduction} 
 In particle accelerators the DLW structure is now mostly distributed and investigated accelerating structure for
acceleration of lightweight particles - electrons in the high energy range, $\beta \approx 1$. 
Naturally realizing advantages of TW mode short RF pulse operation at the S-band and higher frequencies, 
DLW is now dominating normal conducting accelerating structure for these applications.\\
In the L-band frequency range also there are proposals and examples of DLW applications for 
electrons or positrons acceleration, \cite{wanglband}, \cite{keklband}. For lower than L-band 
frequencies DLW application is not effective and there are no examples of applications, \cite{compend}.\\
For lightweight particles acceleration an accelerating system consists from DLW sections with the 
constant period length $d$, matched for particles velocity $\beta=1$.\\
From another side, acceleration of hadrons in linear accelerators also naturally leads to the 
requirements of lower frequencies and SW longer RF pulse operation in medium $\beta$ range. The length 
of period for accelerating structure for hadrons is variable and follows to the growing velocity 
of particles.\\
Comparing the rest energies for protons, muons and electrons, we have $W_{rp}=938.28 MeV, 
W_{rm}=105.66 MeV$ and $ W_{re}=0.511 MeV$, respectively. Acceleration of muons looks more similar to 
protons acceleration. Nevertheless, muons is lighter, as compared to protons, and it looks useful 
to consider solutions, developed for electrons acceleration with DLW, for muons acceleration.\\
The goal of this work is to consider possibility of the DLW structure application for muons 
acceleration in medium and low, as possible, range of particles velocity.  
\section{DLW parameters study}
\begin{figure}[htb] 
\centering
\epsfig{file=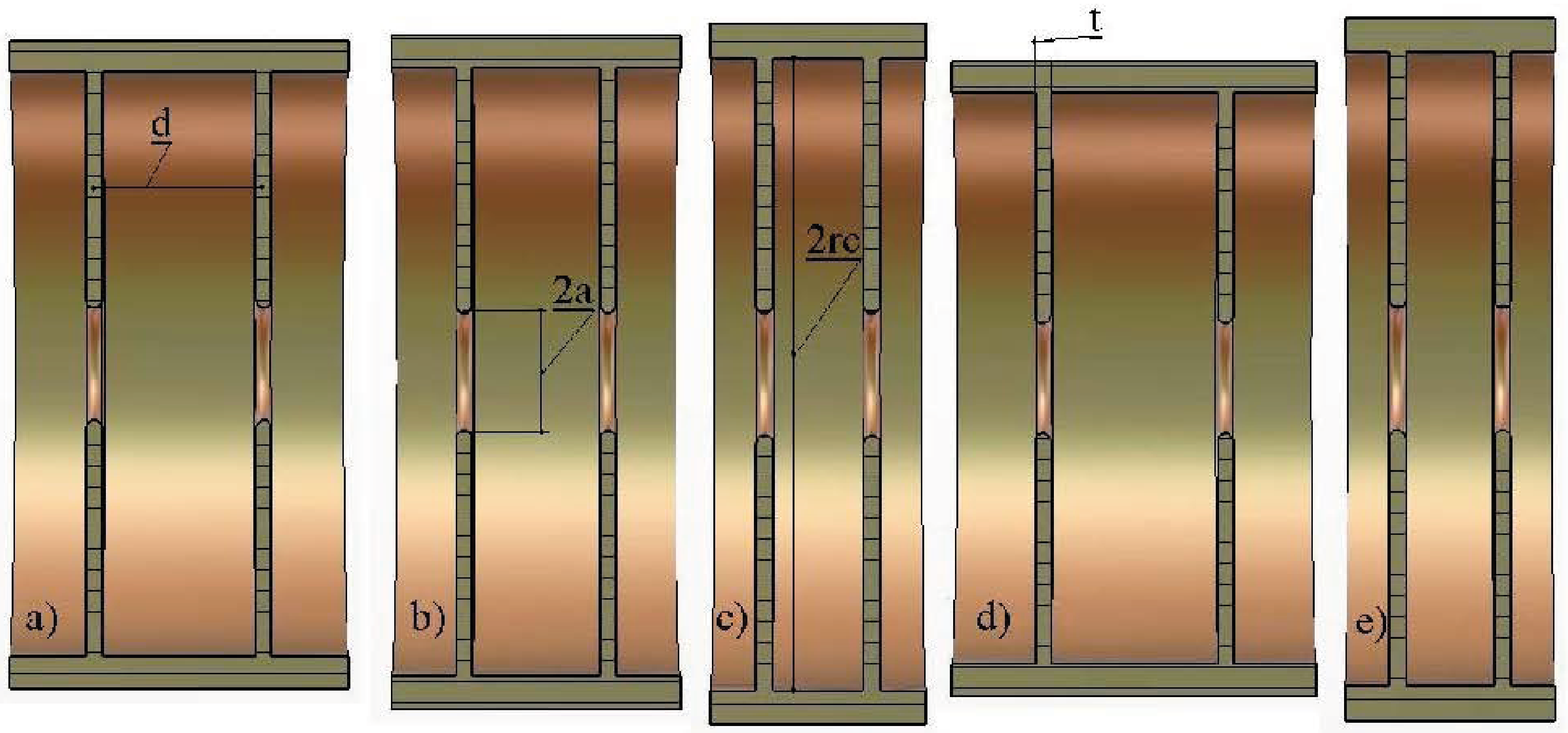, width =145.0mm}
\begin{center}
 Figure 1: The DLW geometry for FW operation $\beta=0.90, \theta=90^o$, (a), $\beta=0.57, \theta=120^o$, (b), 
$\beta=0.41, \theta=120^o$, (c) and BW operation $\beta=0.41, \theta=120^o$, (d), 
$\beta=0.20, \theta=90^o$.
\end{center}
\label{1f}
\end{figure} 
The distribution of the longitudinal electric field $E_z$ in the aperture of the DLW in TW operating mode 
can be presented as the sum over spatial harmonics, see, for example, \cite{lapost}.
\begin{eqnarray}
E_{z}(r,z)=\sum^{n \rightarrow + \infty}_{n \rightarrow - \infty} 
E_{n}I_0( k_{sn} r ) e^{-i k_{zn} z}, \quad k_{zn} =\frac{\theta + 2 n \pi}{d_p}, \quad k^2_{sn}=k^2_{zn}-k^2, \quad
k=\frac{2\pi}{\lambda},
\label{e1}
\end{eqnarray}  
where $E_{n}$ is the amplitude of the $n$-th spatial harmonic, $I_0(k_{sn}r)$ is the modified Bessel 
function, $\lambda$ is the operating wavelength and $\theta$ is the operating phase advance. Usually 
the fundamental, main, spatial harmonic $n=0$ is used for acceleration, due to large $E_0$ value. 
This case DLW has a positive dispersion and operates in Forward Wave (FW) TW mode.\\  
Acceleration with the first nearest spatial harmonic $n=-1$ is possible, see \cite{lapost}, there are 
a lot of papers with proposals, but for $\beta \sim 1$ it loses in RF efficiency, because $|E_0| > |E_{-1}|$. 
Until now just one practically operating facility using DLW with the first harmonic $n=-1$ is known
with an interesting result for particles focusing, \cite{aizaz}. Such case DLW operates in Backward Wave 
(BW) TW mode and has an negative dispersion.\\  
Parameters of the DLW structure were calculated and stored in the data library in the same procedure, 
as described in \cite{library} by using fast and precise 2D finite elements codes \cite{gella}.\\
The considered shapes of DLW cells are shown in Fig. 1 for different $\beta$ and $\theta$ combinations.
The simplest cell shape is considered, because here we investigate DLW possibilities in general. More 
complicated cell shape, similar to considered in \cite{wanglband}, also can be investigated in more 
deep optimization, when solutions with the simple shape are known.\\
The cell shape, shown in Fig. 1, is described by four independent parameters, the aperture radius $a$, 
particle velocity $\beta$, phase advance $\theta$ and iris thickness $t$. The radius of iris rounding 
is accepted as $\frac{t}{2}$ and the cell radius $r_c$ should be adjusted in operating frequency tuning.\\
The length of the cell - the period length $d_p$ is defined for FW or BW operation as 
\begin{equation}
d_p = \frac{\beta \lambda \theta}{2 \pi}\quad or \quad d_p = |\frac{\beta \lambda (\theta -2 \pi)}{2 \pi}|,
\label{e2}
\end{equation}
respectively, assuming $\theta$ value in radians.\\
\begin{figure}[htb] 
\centering
\epsfig{file=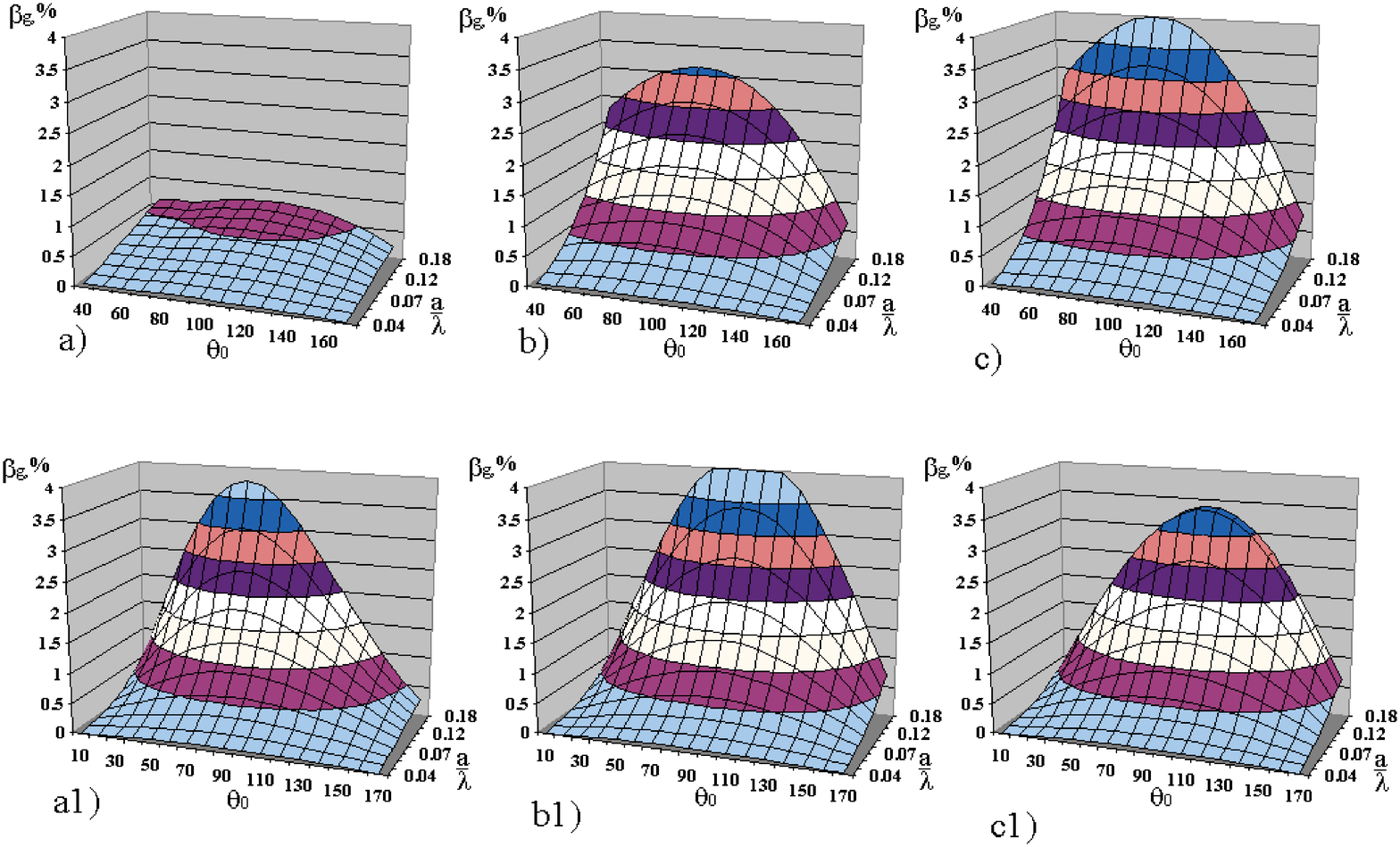, width =155.0mm}
\begin{center}
 Figure 2: The surfaces $|\beta_{g}(\frac{a}{\lambda}|,\theta)$ for $\beta=0.22$ (a,a1), for $\beta=0.57$ (b,b1) and for $\beta=0.92$ (c,c1).
The top line - FW $n=0$ operation, the bottom line - BW $n=-1$ operation, $t=5 mm$.
\end{center}
\label{2f}
\end{figure} 
For operating frequency $f=1296 MHz$ parameters of DLW cells were investigated for independent 
variables in limits:\\
- aperture radius $0.04 \leq \frac{a}{\lambda} \leq 0.19   $ with the step $0.01$;\\
- particles velocity $0.22 \leq \beta \leq 1.0 $ for the forward wave operations and  
$0.04 \leq \beta \leq 1.0 $ for the backward wave operation with the step $0.02$;\\
- operating phase advance $40^o \leq \theta \leq 170.0^o $ for the FW operations and  
$10^o \leq \theta \leq 170.0^o $ for the BW operation with the step $10^o$;\\
- iris thickness $\frac{d_p}{2} \leq t \leq 3 mm$ with $5$ equal steps.
\subsection{Group velocity and quality factor for DLW cells}
The main parameter for DLW TW operation is the group velocity value $\beta_g$. The surfaces 
$\beta_{g}(\frac{a}{\lambda},\theta)$ are shown in the Fig. 2 for different $\beta$ values both for 
FW (the top row in Fig. 2) and BW operation (the bottom row in Fig. 1).\\
Both for FW and BW operation for the constant aperture radius $a =const$ one can see 
sin-like dependence $\beta_g$ on $\theta$ for all $\beta$ values. For the fixed phase advance $\theta$ 
there is a fast rise $\beta_g \sim a^3$ with aperture increasing.\\
\begin{figure}[htb] 
\centering
\epsfig{file=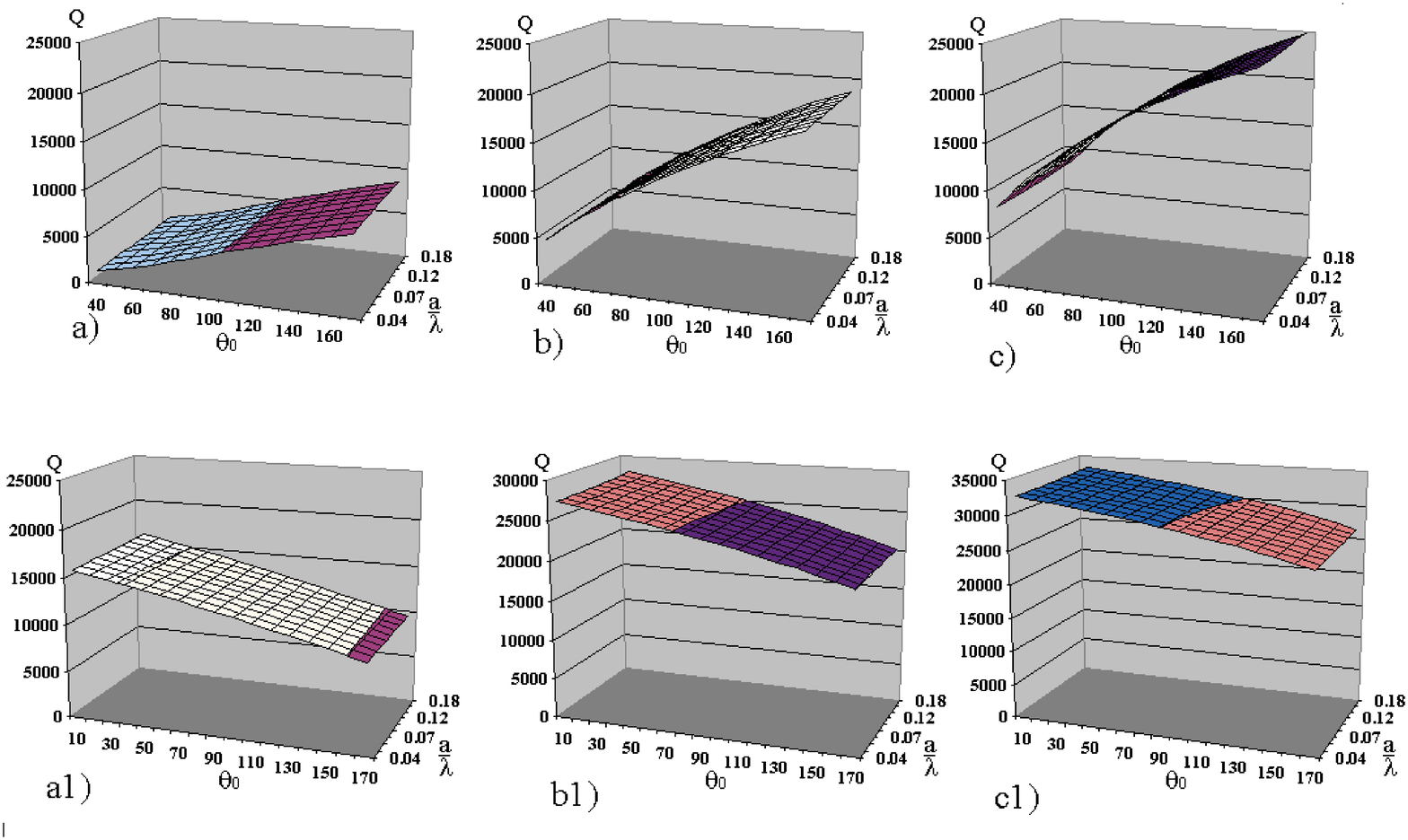, width =155.0mm}
\begin{center}
 Figure 3: The surfaces $Q(\frac{a}{\lambda},\theta)$ for $\beta=0.22$ (a,a1), for $\beta=0.57$ (b,b1) and for $\beta=0.92$ (c,c1).
The top line - FW $n=0$ operation, the bottom line - BW $n=-1$ operation, $t=5 mm$.
\end{center}
\label{3f}
\end{figure}
Another important parameter of DLW cells is the quality factor $Q$ and the surfaces $Q(\frac{a}{\lambda},\theta)$ 
are shown in Fig. 3 for different $\beta$ values both for 
FW (the top row in Fig. 3) and BW operation (the bottom row in Fig. 3).\\ 
The structure operates in $TM_{01}$ - like mode and the main parameter, which defines $Q$ value is the 
ratio of the cell length to the cell radius. As one can see from the surfaces in Fig. 3, there are 
no essential $Q$ dependence on the aperture radius $a$. For lower $\beta$ values quality factor decreases, 
especially for FW operation. With $\theta$ decreasing the cell length decreases for 
FW operation and increases for BW one. It explains the opposite slope of surfaces 
in Fig. 2 for forward and backward waves. For the first synchronous harmonic $n=-1$ the cell length 
is all time larger, than for synchronous fundamental harmonic $n=0$ and quality factor is higher all 
time.
It results in lower wave attenuation for BW operation.
\subsection{Main relations. Frequency scaling and $\beta$ limitations}
To point out the main relations and frequency scaling for DLW parameters we will write below well 
known formulas. For this subject one can see also the classical papers with discussions about DLW parameters 
and relations  \cite{lapost}.\\      
For TW operation the magnitude of the $n$-th accelerating harmonic $E_{n}$ is related with the flux of RF 
power in the traveling wave $P_t$ as 
\begin{equation}
\frac{E_n}{\sqrt{P_t}} = \sqrt{\frac{2 \pi Z_{en}}{ \lambda |\beta_g| Q}},\quad [\frac{V}{\sqrt{Wt}}],
\label{e3}
\end{equation}
where $Z_{en}$ is the value of the effective shunt impedance per unit of length for the $n$-th harmonic, 
\begin{equation}
Z_{en}= \frac{|\int_{d_p} E_z(0,z) e ^{ik_{zn} z} dz|^2}{P_s d_p}, \quad Q =\frac{4 \pi f W_e}{P_s},
\quad W_e=\frac{\epsilon_0}{2} \int_V |E|^2 dV,
\label{e4}
\end{equation}
$P_s$ is the power of RF losses in the surface and taking into account that in the traveling wave the RF energy is stored 
simultaneously both in electric and magnetic field,
$W=W_m+W_e=2W_e$. Substituting relations from (\ref{e4}) into (\ref{e3}), we find 
\begin{equation}
\frac{E_n}{\sqrt{P_t}} = \sqrt{\frac{ |\int_{d_p} E_z(0,z) e ^{ik_{zn} z} dz|^2}{ d_p c |\beta_g| 
\epsilon_0 \int_V |E|^2 dV }}. 
\label{e5}
\end{equation}
The structure operates in $TM_{01}$-like mode and electric field distribution along radius is shown 
in Fig. 4a for different $\beta$ at the distance $\frac{d_p}{3}$ from the iris center. The maximal 
field value $E_{max}$ takes place at the iris tips. In the beam aperture $\frac{r}{\lambda} \leq 0.1 $ 
one can see electric field decay to the axis due to natural harmonics attenuation. To estimate 
the integral over cell volume in (\ref{e5}), we can approximate the field distribution in the cell similar 
to a simple cylindrical cavity $E_z(r) \sim E_{max} J_0(kr)$, due to relatively small volume 
of beam aperture, neglecting field attenuation in this region. With such approximation we have  
\begin{equation}
\int_V |E|^2 dV  \approx 2 \pi d_p E_{max} ^2 \int_0 ^{r_{c}} J_0^2(kr) r dr  
=  \pi d_p E_{max} ^2 r^2_c J^2_1(k r_c)  =  \pi d E_{max} ^2 \frac{\nu_{01}^2 
 J^2_1(\nu_{01})}{k^2},  
\label{e6}
\end{equation}
where $\nu_{01}=k r_c =2.4048$ is the first root $J_0(x)=0$, and $J_0(kr)$ is the Bessel function.\\  
\begin{figure}[htb] 
\centering
\epsfig{file=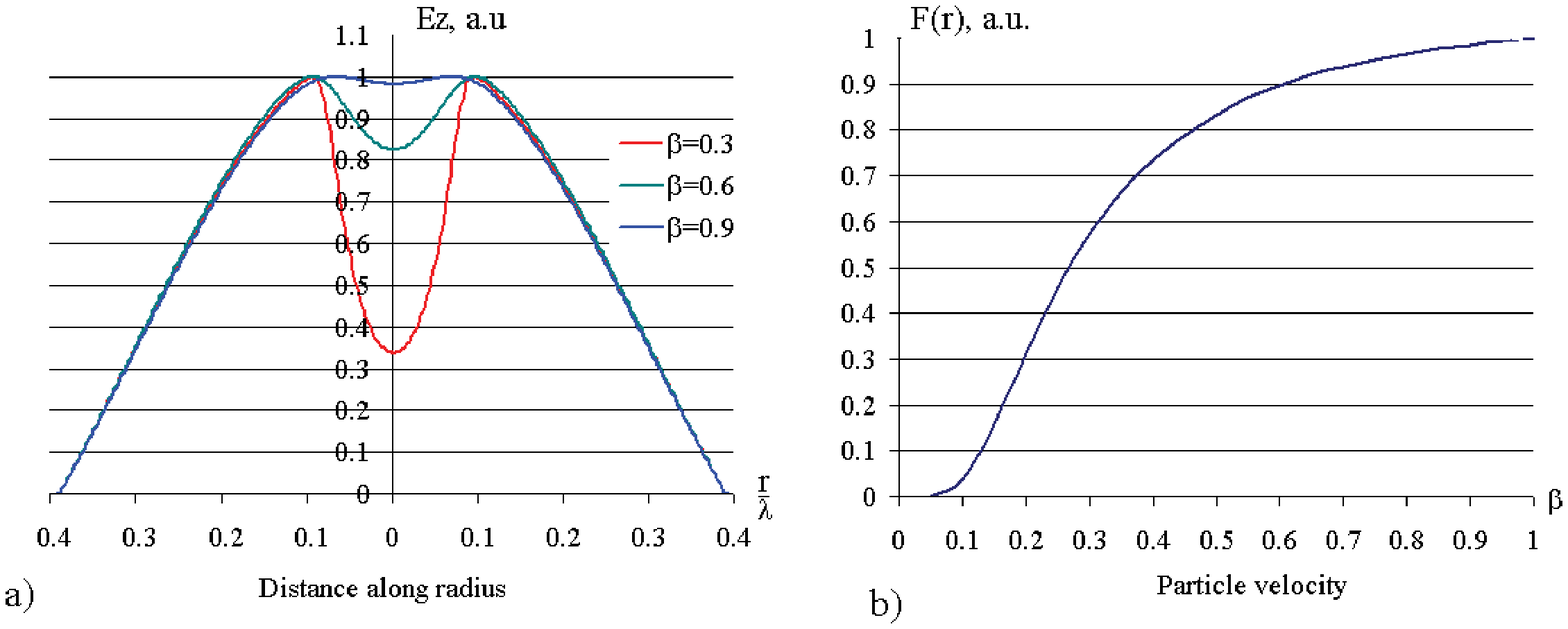, width =145.0mm}
\begin{center}
 Figure 4: The electric field distribution along radius for different $\beta$ values at the distance 
$\frac{d}{3}$ from the iris center, (a) and the plot of the function 
$I_0( \frac{2 \pi a \sqrt{ 1 -\beta^2}}{\beta \lambda}  )$ for $\frac{a}{\lambda}=0.08$, (b).
\end{center}
\label{4f}
\end{figure}
The integral over DLW axis in (\ref{e4}) is the part of the RF voltage along axis due to main spatial 
harmonic. From (\ref{e1}) and (\ref{e2}) we can estimate the attenuation of the main harmonic 
from the iris aperture to the axis as 
\begin{equation}
|\int_{d_p} E_z(0,z) e ^{ik_{zn} z} dz|^2 =(E_0 d_p)^2 
\approx (E_{max} d_p I_0( \frac{2 \pi a \sqrt{ 1 -\beta^2}}{\beta \lambda}  )^2.
\label{e7}
\end{equation}
The plot of the function $I_0( \frac{2 \pi a \sqrt{ 1 -\beta^2}}{\beta \lambda}  )$ for $\frac{a}{\lambda}=0.08$ 
is shown in the Fig. 4b. In this approximation we neglect the contribution of higher spatial harmonics 
at the axis, which attenuate from the iris aperture faster than fundamental. This contribution can 
be taken into account by transit time factor $T$, which is $T \sim 0.8, \beta \sim 1, \theta \sim 120^o$ 
and $T \sim 1$ for lower value of the $\beta \cdot \theta$ product.\\
Substituting results of (\ref{e7}) and (\ref{e6}) into (\ref{e5}), one can get
\begin{equation}
\frac{E_0}{\sqrt{P_t}} \approx 
\frac{ I_0( \frac{2 \pi a \sqrt{ 1 -\beta^2}}{\beta \lambda}  )}
{\lambda \nu_{01} J_1(\nu_{01})} \sqrt{\frac{2 \pi Z_0}{|\beta_g|}} \approx 
100 \frac{I_0( \frac{2 \pi a \sqrt{ 1 -\beta^2}}{\beta \lambda}  )}{\lambda \sqrt{|\beta_g|}}, \quad Z_0 = \sqrt{\frac{\mu_0}{\epsilon_0}}. 
\label{e8}
\end{equation} 
This approximated relation allows us estimate dependences of DLW parameters for FW operation 
on operating frequency $f=\frac{c}{\lambda}$, particle velocity $\beta$ and wave group velocity 
$\beta_g$. For example, with the input RF power $P_t = 1 MWt$, group velocity $\beta_g=0.01$ and 
$\lambda =0.23 m$ one can expect $E_0 \sim 4 \frac{MV}{m}$. As it will be shown later, see Section 
4, estimation (\ref{e8}) provides $\sim 20\%$ overestimation in $E_0$ value, because transit time 
value $T$ is not included in (\ref{e8}). Including $T$ value in estimation, we have very good 
preliminary estimation for possible value of accelerating gradient.\\ 
Another important DLW parameter is the attenuation constant 
\begin{equation}
\alpha =\frac{\pi}{\lambda |\beta_g| Q},\quad [\frac{Np}{m}].
\label{e9}
\end{equation} 
\subsubsection{Frequency scaling}
As one can see from (\ref{e8}), the ratio $\frac{E_0}{\sqrt{P_t}} $ scales as $f^{-1}$, limiting 
DLW application for low frequencies. Comparing the S-band and the L-band ranges, we have ratio $\approx 2.2$. 
To get the same accelerating gradient with the same group velocity $\beta_g \sim (1 \div 2) \% $ will 
require $\sim 5 $ times more RF power in the L-band range, as compared to the S-band range. 
For DLW application in the L-band range the lower $\beta_g$, as compared to the S-band range, should 
be used and low proposed values  $\beta_g \sim (0.37 \div 0.12)\%$ and  $\beta_g \sim (0.67 \div 0.20)\%$ are considered 
in \cite{wanglband}. In the DLW section \cite{keklband} also $\beta_g \sim (0.61 \div 0.39)\%$ is realized.\\
Another DLW parameter, attenuation (\ref{e9}), scales as $f^{-\frac{3}{2}}$. Instead of $Q$ reduces with 
$\beta$ reduction, see Fig. 3, even with the reduced 
group velocity $\beta_g$ we can not expect the same $\alpha$ value as for S-band application.
\subsubsection{Limitations from particle velocity}
With particles velocity $\beta$ decreasing the natural field decay, see (\ref{e7}), Fig. 4b, decreases further 
DLW efficiency. The field decay for low $\beta$ is the common property for all accelerating structures. 
For DLW this problem is more sharp, because the aperture radius $a$ should be defined to get required 
group velocity value $\beta_g$. The sequence of this field decay is a strong increasing of 
$\frac{E_{smax}}{E_0}$ ratio, where $E_{smax}$ is the maximal value of electric filed at the cell surface.
The surfaces $\frac{E_{smax}}{E_0}(\frac{a}{\lambda},\theta)$ are shown in the Fig. 5 for different 
$\beta$ values both for FW (the top row in Fig. 5) and BW DLW operation (the bottom row in Fig. 5).
One can see $\frac{E_{smax}}{E_0}$ rise with $\beta$ decreasing for FW operation and high 
$\frac{E_{smax}}{E_0}$ value for DLW BW operation.
\begin{figure}[htb] 
\centering
\epsfig{file=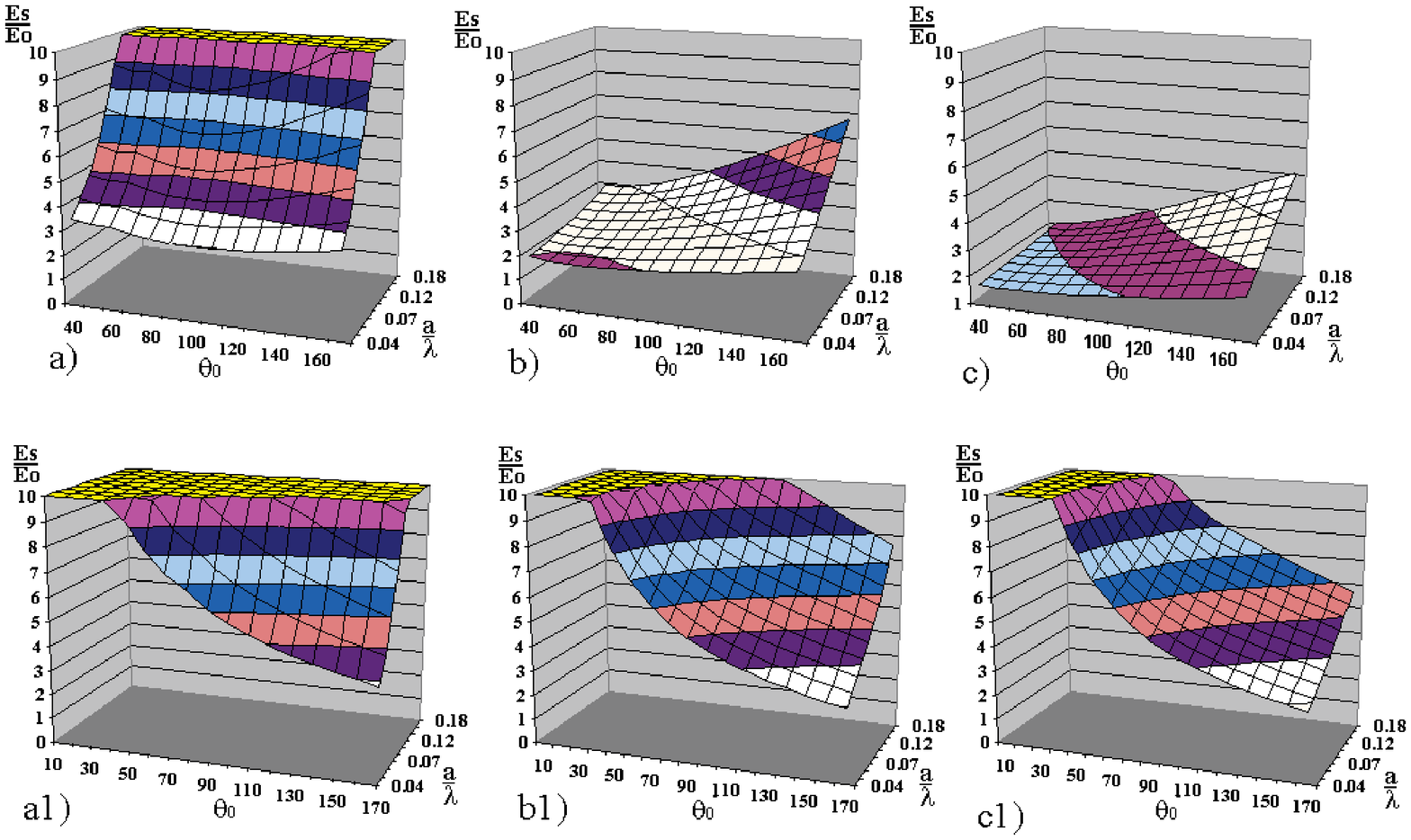, width =155.0mm}
\begin{center}
 Figure 5: The surfaces $\frac{E_{smax}}{E_0}(\frac{a}{\lambda},\theta)$ for $\beta=0.22$ (a,a1), 
for $\beta=0.57$ (b,b1) and for $\beta=0.92$ (c,c1).
The top line - FW $n=0$ operation, the bottom line - BW $n=-1$ operation, $t=5 mm$.
\end{center}
\label{5f}
\end{figure}  
\subsection{The optimal iris thickness}
In the reference \cite{keklband} relatively thick iris $t=10 mm$ is applied. Probably it come 
from direct scaling of the DLW cell dimensions from the S-band to the L-band range. In Fig. 6a 
are shown the plots of the maximal $\frac{E_0}{\sqrt{P_t}}$ values assuming $P_t=1 MW, \beta_g=0.01$ for different 
iris thickness $t$ values. To get points at the plots, for each $\beta$ and $t$ values there was a scan 
over $40^o \leq \theta \leq 170^o$ to find the maximal $\frac{E_0}{\sqrt{P_t}}$ value. 
It means, that along each curve in Fig. 6a $\theta$ changes with $\beta$ and $t$. Corresponding 
plots of $\frac{E_{smax}}{E_0}$ ratio are shown in Fig. 6b.\\
\begin{figure}[htb] 
\centering
\epsfig{file=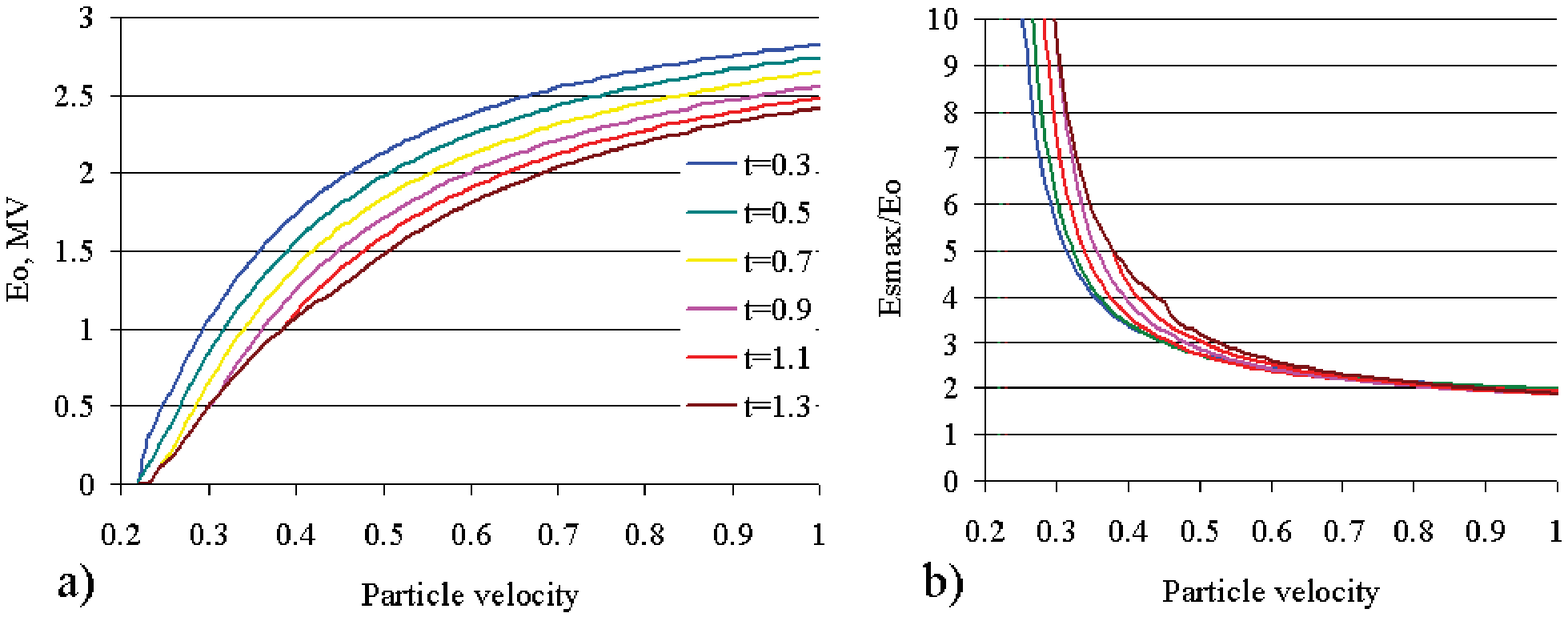, width =155.0mm}
\begin{center}
 Figure 6: The plots of the maximal $\frac{E_0}{\sqrt{P_t}}$ values assuming $P_t=1 MW, \beta_g=0.01$ (a) 
and corresponding plots of $\frac{E_{smax}}{E_0}$ ratio (b).
\end{center}
\label{6f}
\end{figure}  
As one can see from Fig. 6a, for the thin iris we can obtain higher $\frac{E_0}{\sqrt{P_t}}$ value 
without practical increasing of the maximal field at the surface, Fig. 6b. With thin iris we obtain 
the required value of group velocity $\beta_g$ with smaller aperture radius. It results in 
$\frac{E_0}{\sqrt{P_t}}$ increasing and partially compensates $\frac{E_{smax}}{E_0}$ increasing 
due to iris thickness decreasing. Also slightly increases the quality factor of the cell and 
decreases attenuation. It allows to get the higher total energy gain for DLW section.\\
For the L-band application the iris thickness should be minimal, as possible, from mechanical rigidity 
and heat transfer requirements. In the DLW operations with a short RF pulse $\sim 5 \mu s$ the average 
heat loading is negligible and heat transfer limitations (or iris cooling) are not required.\\
As for mechanical rigidity, iris thickness $t \approx 5 mm$ looks sufficient. May be iris deformations
during DLW section brazing. But such deformations take place all time and RF tuning after brazing is 
required all time to compensate the effect of such deformations. Moreover, in the periodical DLW 
structure assuming an uniform heating during brazing we can expect  similar deformations for adjacent 
irises, which will lead to similar changes of frequencies for adjacent cells.\\ 
For DLW application in the L-band range the reduced iris thickness is preferable to get the 
maximal accelerating gradient and the total energy gain of DLW section. 
\subsection{The maximal accelerating gradient and the maximal energy gain for FW 
constant gradient operation}
Because the cell length for intermediate particles acceleration should be adjusted with growing 
particle velocity, all cells in DLW section have different length and constant impedance mode 
can not be supported. 
The reasonable mode for DLW FW operation is the constant gradient mode, 
which is mostly used for $\beta=1$ case. In this way for DLW section two values should be defined - 
the input and the output group velocity, $\beta_{g_{in}} > \beta_{g_{ou}}$.\\  
\begin{figure}[htb] 
\centering
\epsfig{file=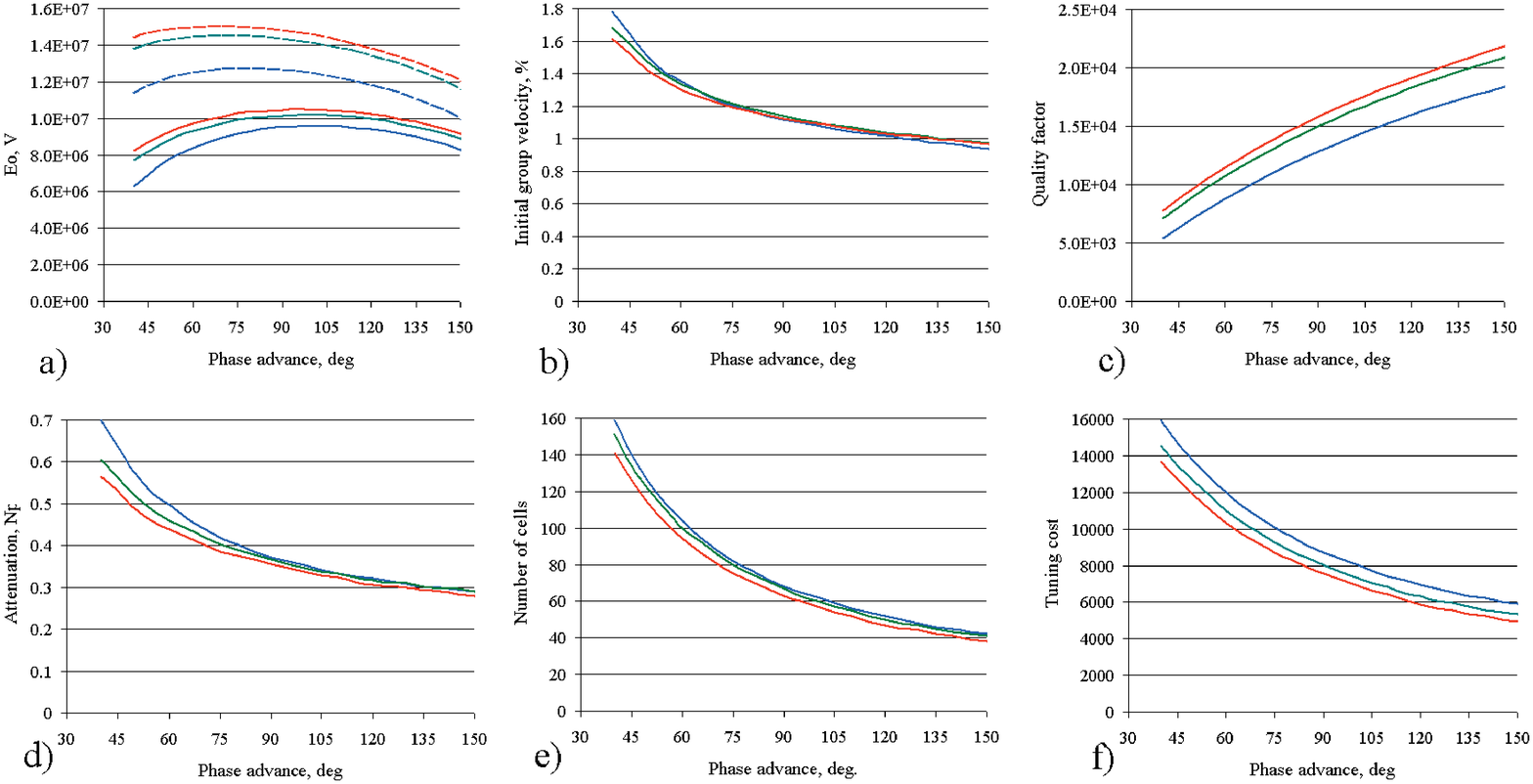, width =160.0mm}
\begin{center}
 Figure 7: Some data for constant gradient DLW sections (see text for explanations). 
The maximal possible accelerating gradient (dotted curves) and realized accelerating gradient 
(solid curves) (a), the initial group velocity  $\beta_{g_{in}}$ assuming $\beta_{g_{ou}} =0.6\%$ (b), 
the quality factor for DLW cells (c), the total attenuation for DLW sections (d), the number 
of cells in the DLW sections (e) and the 'tuning costs' for DLW sections.
\end{center}
\label{7f}
\end{figure}  
In the Fig. 7 some parameters of the constant gradient sections in the dependence on operating 
phase advance $\theta$ are illustrated for three examples - the DLW section with the output group 
velocity $\beta_{g_{ou}} =0.6\%$ and:\\
 - input particle velocity $\beta =0.56197$, the length $\approx 2.5 m$ - blue curves;\\
 - input particle velocity $\beta =0.79098$, the length $\approx 3.17 m$ - green curves;\\
 - input particle velocity $\beta =0.89447$, the length $\approx 3.3  m$ - red curves.\\
These examples illustrate the DLW sections parameters with variable cell length for FW constant 
gradient operation in the middle and the high $\beta$ range. Important point is that for all 
three examples we assume the assume the same output, the minimal for the section, 
group velocity $\beta_{g_{ou}} =0.6\%$ and the same input RF power of $18 MW$. Also the 
synchronous phase for the particles is suggested as $\phi_s=\frac{\pi}{6}$ for the first 
example and as $\phi_s=\frac{\pi}{18}$ for the second and the third.\\
With the dotted curves in the Fig. 7a are shown the plots of the maximal $E_0$ value, according 
(\ref{e3}), for different $\theta$ values and $\beta_g=0.6\%$. One can see enough high $E_0 \sim 
(12 \div 15) MV/m$ values with the maximums near $\theta \approx 90^o$ for $\beta \sim 0.5$, shifting 
to $\theta \approx (60^o \div 75^o)$ for $\beta \sim 0.9$. So, we see $E_0$ maximum at low 
operating phase advance. But results for constant gradient sections shows lower results with the 
maximal values for total energy gain at higher phase advance $\theta \approx (90^o \div 120^o)$. 
One can see clearly from Fig. 7a, that the maximal $E_0$ values at the solid curves are shifted to higher
$\theta$ values with respect dotted lines.\\ 
If we select for a DLW section a low phase advance $\theta \leq 90^o$, going to use the maximal 
$E_0$ value according dotted curves in Fig. 7a, it means shorter cells, (\ref{e2}), and, 
as the sequence, lower quality factor, Fig. 7c, and higher attenuation constant (\ref{e9}), as one can see 
from Fig. 7b. Designing DLW section with the fixed $\beta_{g_{ou}}$ for lower $\theta \leq 90^o$ values we 
need in larger $\beta_{g_{in}}$ values, as compared to $\theta \geq 90^o$. To keep the gradient constant, 
we have to compensate larger wave attenuation at lower $\theta$ by faster $\beta_g$ decreasing. 
As one can see from the Fig. 7b, the difference $\beta_{g_{in}} - \beta_{g_{ou}}$ decreases with 
$\theta$ increasing. But, if we take higher $\beta_{g_{in}}$ values, we reduce, according (\ref{e3}), 
realized $E_0$ value, in comparison with possible for lower $\beta_g$, in the section beginning. 
In the section end the wave is already attenuated and RF power is lower, as compared to the input value. 
So, as one can see from Fig. 7, the optimal, for the maximal energy gain at the DLW section with 
the constant gradient, values of the phase advance $\theta$ are shifted to the region $90^o \leq 
\theta \leq 120^o$. The shift depends on the total attenuation $\tau$ in the section. This maximum is 
smooth enough and the relative difference in the total energy gain $\delta W$ is not more than $3 \%$ 
between $\theta=90^o$ and $\theta=120^o$ for small attenuation, $\tau \sim 0.25 Np$, but with essential 
difference in the number of cells in the section, Fig. 7e. Such case we have to discuss more another 
practical issue - DLW section tuning.
\subsection{Tuning efforts}  
Tuning of cells frequencies for DLW section is all time required after section brazing to provide 
 the required distribution of the phase for traveling wave. The errors in the phase of the traveling wave $\delta \theta$ are 
connected with the error $\delta f_c $ in the cell frequency $ f_c $ as 
\begin{equation}
\frac{\delta \theta}{\theta} = \frac{ \delta f_c}{ \beta_g f_c}.
\label{e10}
\end{equation} 
There are methods for tuning very long DLW sections after brazing, see for example \cite{tuning}. 
As one can see from (\ref{e9}), more rigid tolerances for deviations of cells frequencies are 
required for the same phase deviations in DLW sections with lower $\beta_g$ values and tuning is more 
difficult. At least more iterations in cells tuning can be required. To have something quantitative to 
compare possible efforts in DLW sections tuning with different number of cells and different $\beta_g$ values,
we suggest such parameter - 'tuning costs' $T_c$ as 
\begin{equation}
T_c = \sum_{i=1}^{N} \frac{1}{ \beta_{g_i} }.
\label{e11}
\end{equation}  
where $N$ is the number of DLW cells in the section and $\beta_{g_i}$ is the value of group velocity 
for the $i$-th cell. This parameter has no physical sense and is introduced as an attempt for 
some numerical comparison. The value of this parameters grows with the number of cells in the section 
and reflect more rigid tolerances for cells frequencies according to (\ref{e10}).\\
Considering this parameter dependence on $\theta$ value, Fig. 7f, one can see $T_c$ decreasing 
for $\theta$ increasing. For the fixed length of the DLW section and the fixed $\beta_{g_{ou}}$ value 
with $\theta$ increasing the reduction of the number of cells in the section, Fig. 7e, is more fast 
than reduction of the average group velocity in the section. While the difference in the energy gain 
$\delta W$ between DLW sections with $\theta=90^o$ and $\theta=120^o$ is small, ($ \sim (1 \div 3)\%$), 
the difference in $T_c$ is more sufficient, ($ \sim (20 \div 30)\%$). It may be an argument for a choice 
higher $\theta \sim 120^o$, as compared to calculated optimal $\theta \sim 90^o$ - the defeat in 
energy gain is at the level of precision for simulations and measurements, but the profit in 
the reduction of efforts for section tuning is quite visible.
\subsection{RF efficiency and power requirements for TW and SW operation} 
As one can conclude from plots in the Fig. 7, for the DLW section with input particle velocity $\beta_{in} =0.56197$ and the 
length $\approx 2.5 m$ for the input RF power $18 MW$ we can expect accelerating gradient $E_0 \approx 9.4 \frac{MV}{m}$ 
and energy gain $W \approx 20.8 MeV$. For another section with $\beta_{in} =0.89447$ and the 
length $\approx 3.3 m$ we can expect $E_0 \approx 10.22 \frac{MV}{m}$ 
and $W \approx 33.2 MeV$. The filling time for traveling wave DLW structure with the length $\approx 3 m$ and group 
velocity $\beta_g \sim 1\%$ is $\tau_f \approx 1 \mu s$.\\
For the L-band SW normal conducting structures, which normally have drift tubes for higher RF efficiency, 
the values of accelerating gradient $E_0 \approx 10 \frac{MV}{m}$ are high enough. Because requirements of high 
effective shunt impedance $Z_e^{(SW)}$ and high $E_0$ are contradictory for structures with drift tubes, we 
can expect $Z_e^{(SW)} \sim 35 \frac{MOm}{m}$ for $\beta_{in}  \sim 0.55$ and $Z_e^{(SW)} \sim 45 \frac{MOm}{m}$ for 
$\beta_{in}  \sim 0.9$. \\
To develop the same energy gain as for two TW DLW sections, mentioned before, we will need pulse RF power $\approx 
6.3 MW$ and $\approx 7.7 MW$ with SW structures. The SW operating mode with an appropriate accelerating structure is 
$\approx 3 $ times more economic in the required pulse RF power. \\   
The typical value of quality factor for SW structure in the L-band range is expected as $\approx 22 000$. It 
corresponds to field rise time $\tau_r = \frac{Q}{2 \pi f } \approx 2.7 \mu s$. During transient, 
which takes at least $ \geq 3 \tau_r \approx 10 \mu s$, RF power reflects from SW structure and should 
be dissipated in RF load by using circulators or $3dB$ hybrids.\\
Comparing TW and SW structures, we came to requirements - either short $\approx (4 \div 6) \mu s$ RF pulse with the 
pulse RF power $\sim 20 MW$ or much longer RF pulse $\approx 20 \mu s$ but with lower RF pulse power $\sim 7 MW$. Both 
options can be realized. The parameters of RF sources - klystrons and corresponding parameters of modulators are limited
 essentially by the average RF power, which is approximately the same for both considered options.\\ 
Requiring higher pulse RF power, TW operating mode requires approximately the same RF energy due 
to longer pulse length for SW mode.\\
For high operating frequency the klystrons with higher RF pulse power, at the expense of short RF 
pulse, are known. For the L-band TW application with DLW the klystron with the pulse RF power of 
$40 MW$ and the pulse width of $4 \mu s$ is assumed, \cite{keklband}. Moreover, at least for a high operating frequencies 
(the S band and higher), the technique for pulse compression is known and developed, see, for example, 
\cite{pulse}. It allows to get approximately twice higher RF pulse power, as the output pulse RF 
power of the klystron. The same technique is assumed for RF system with DLW structure both in 
\cite{keklband} and in \cite{muon}.\\
Considering examples of possible accelerating systems with DLW, we also will assume that the 
pulse RF power $\geq 80 MW$ cam be obtained from a single RF source. 
\section{Comparison with the known DLW L-band examples} 
Before structure design, our data library with DLW parameters and codes for DLW sections design were 
compared in final results with similar results, published by another authors for L-band DLW sections. The 
mostly representative data for L-band sections with the same cell shape are given in \cite{keklband} 
and results are reproduced in the second column of the Table 1 as 'Reference'. The results in \cite{keklband} 
are given for the input RF power of $15 MW$. For the same RF power in the third column 'Comparison' 
are given similar parameters, obtained with our data library. Comparing numbers in the second and the 
third column in Table 1, one can see small differences, which are at the relative level of $(1 \div 3)\%$. 
It can be explained by precision of calculations for initial data with different 2D software. The mostly 
distributed 2D code with sufficient possibilities is well known SuperFish \cite{lanlcode}. Because the 
codes set \cite{gella} use higher order approximations of the field components, long time experience 
shows that results of simulations with \cite{gella} are more precise. Anyhow, the difference in results 
between 'Reference' and 'Comparison' is very small.\\   
\begin{table}[htb]   
\begin{center}
\centering{Table 1: The DLW data library results calibration at KEKB L-band structure.\\}
\begin{tabular}{|l|c|c|c|c|c|c|c|c|c|c|c|c|c|c|c|c|}
\hline
                 & Reference        & Comparison         & Option 1            & Option 2             \\
\hline
$\theta$, deg.   & 120              & 120                & 120                 & 90                   \\
\hline
$t, mm$          & 10               & 10                 &  5                  &  5                   \\
\hline
Gradient, MV/m   & 12               & 11.85              & 12.57               & 12.91                \\
\hline
$ 2a, mm$        &$39.4 \div 35.0$  & $39.76 \div 35.21$ & $34.77 \div 30.334$ & $33.992 \div 29.014$ \\
\hline
$\beta_g, \%$    &$0.61 \div 0.39$  &$0.634 \div 0.39$   &  $0.629 \div 0.39$  & $0.683 \div 0.39$    \\
\hline
$\tau, Np$       & 0.261            & 0.2626             &  0.2692             &  0.3129              \\
\hline
$Q$              & $\sim 20000$     & $\sim 20370$       & $\sim 21190$        & $\sim 17600$         \\
\hline
Length, m        & 2                &  2                 &  2                  &  2.02                \\       
\hline
N cells          &                  &  26                &  26                 &  35                \\       
\hline
$Tuning, T_c$    &                  &  5105              &  5133               &  6603                \\       
\hline
\end{tabular}
\label{1t}
\end{center}
\end{table} 
Further we apply our conclusions, which were done in above consideration of DLW parameters. 
In the fourth column 'Option 1' the iris thickness is reduced to $5 mm$ and phase advance is the same as 
before - $\theta = 120^o$. It results in increasing of accelerating gradient and total energy gain 
at $\approx 6\%$ with associated aperture radius reduction $a$ at $2.5 mm, \sim 16 \%$. If aperture 
radius decreasing is tolerable, we can have a sufficient gain in $E_0$ value.
Here we test also our parameter tuning efforts, $T_c$, (\ref{e11}). As we can see from the Table 1,
tuning efforts $T_c$ are practically the same both for $t=10 mm$ and for $t=5 mm$.\\ 
Another improvement can be obtained by choosing the phase advance $\theta =90^o$, 'Option 2' in the 
fifth column. There will be an additional $E_0$ improvement in $\sim 3\%$, as compared to 'Option 1',
but number of cells in the section increases from $26$ to $35$. This choice $\theta =90^o$ may be not 
so evident from practical reasons, because the increasing of cells number results to increasing in tuning efforts $T_c$ 
at $\approx 30\%$.    
\section{Forward and backward wave operation}
\begin{figure}[htb] 
\centering
\epsfig{file=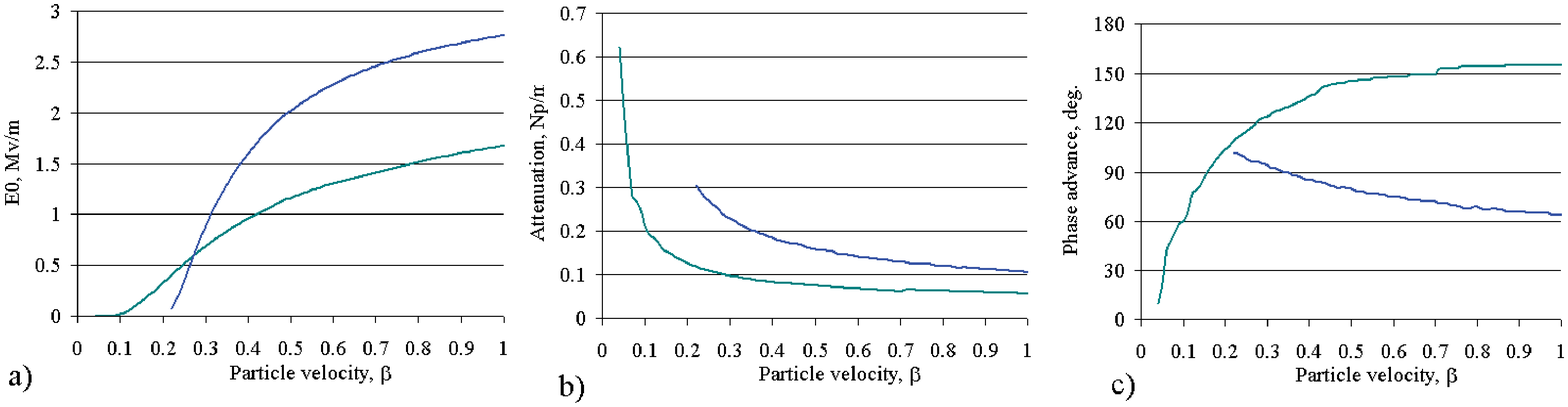, width =160.0mm}
\begin{center}
 Figure 8: The plots of the maximal $\frac{E_n}{\sqrt{P_t}}$ values assuming $P_t=1 MW, \beta_g=0.01$ (a) 
and corresponding plots of attenuation $\alpha$ (b) and optimal phase advance $\theta$, (c).
The blue curves correspond to FW $n=0$ operation, the green curves are for BW $n=-1$ operation, $t=5 mm$.
\end{center}
\label{8f}
\end{figure}  
Normally the DLW structure is used for high energy application in FW mode with the fundamental $n=0$ 
synchronous spatial harmonic because this harmonic has the maximal amplitude in the expansion 
(\ref{e1}) and the value of shunt impedance $Z_{e{0}}$ is higher. Nevertheless, the possibility 
of DLW application with $n=-1$ synchronous is not prohibited, is mentioned in classical books 
\cite{lapost} and realized in \cite{aizaz} for $\beta=1$.\\  
The comparison of the maximal $E_n$ values for FW and BW operations is shown in Fig. 8a. In the region 
$\beta \leq 0.28$ BW operation provides higher value of acceleration gradient. Another BW preference 
for low $\beta$ region is in the smaller value of attenuation $\alpha$, Fig. 8b, due to longer cells, 
(\ref{e2}), and, as the sequence, higher value of quality factor, see Fig. 3.\\
There is additional advantage, leading to higher energy gain at DLW sections in low energy region. 
As one can see from Fig. 8a, both for BW and FW operation the maximal $E_n$ value increases with 
$\beta$ increasing. In FW sections, $\beta_g >0$ RF input is placed in the section beginning. 
The maximal traveling power is in the section part with lower $\beta$. In this part the maximal possible 
$E_0$ value is lower, than for the region with higher $\beta$ near the section end, and wave attenuation 
is higher, see Fig. 8a, 8b. To the end of the section, where higher $E_0$ values can be obtained, 
the wave comes already attenuated. For BW operation, $\beta_g < 0$, RF input is places in the section 
higher $\beta$ end, where the maximal possible gradient can be achieved and attenuation is lower. 
This particularity leads to higher energy gain for BW DLW section in higher region of particles velocity, 
$\beta \leq (0.4 \div 0.5)$.\\
To extend accelerating DLW structure  to low $\beta$ region, the backward wave operation 
mode is more effective.\\
The plots of the optimal $\theta$ value to get the maximal accelerating gradient are shown in Fig. 8c.
The DLW cell length in BW mode is not so critical to $\theta$ value choice as for FW mode, (\ref{e2}). It 
relaxes a technical limitations for DLW cells in low $\beta$ region.\\
There is an additional comment to the plots in Fig. 8. The dependence $E_0 (\beta)$ in Fig. 8a for FW operation 
is obtained in the treatment of numerical data library. One can see, that the plot $E_0 (\beta)$ in Fig. 8a 
is similar to the plot of the function $I_0( \frac{2 \pi a \sqrt{ 1 -\beta^2}}{\beta \lambda}  )$ in Fig. 4b. 
Comparing numbers, one can conclude - the approximation (\ref{e8}) provides good qualitative 
and quantitative estimation for DLW accelerating gradient in FW mode.   
\section{Examples of accelerating structure design for muons acceleration}
In this section we will consider several examples of DLW accelerating structure for muons acceleration. 
Let us discuss, at first, the design approach and the reasonable safety margins.
\subsection{Design approach for DLW sections choice for muons acceleration}
Because muons are not stable particles, it should be accelerated as fast, as possible. 
And the main guideline for design of DLW sections is the maximal possible accelerating gradient, 
or the maximal energy gain both for each DLW section and for the total DLW accelerating system.\\ 
Parameters of the DLW sections can be considered with different initial assumptions. The main parameter 
for the choice is the group velocity $\beta_g$ in the section. For higher accelerating gradient 
as low as possible, $\beta_g$ value is desirable. To overlap the wide region of particle velocities, 
DLW accelerating structure should use BW operating mode for low $\beta$ region, and usual FW operating 
mode for moderate and high $\beta$ region. Taking into account reasons of practical realizations and 
sections tuning, at first we suggest the same minimal group velocity in all DLW sections, 
in the DLW accelerating structure, both for BW and for FW operation.\\
Because the cell length follows to the growing particle velocity, all cells in the section are slightly 
different in dimensions. This case we can not take advantage for cost reduction in production from 
the constant impedance section option, which is possible for DLW with $\beta=1$. But we can consider 
the option of the constant group velocity $\beta_g$ along the section and the option of the constant 
accelerating gradient $E_0$.\\
In the beginning of the accelerating system, $\beta \sim 0.2$, where BW operation is more effective, 
due to the fast change of DLW parameters with rising $\beta$, see plots in Fig. 8, the constant 
gradient choice is not reasonable. In DLW sections with BW operation RF input is placed at the 
section end, in the place of higher $\beta$. As one can see from the plots in Fig. 8a, if we will 
try to keep $E_0 = const$ from the section end to the section beginning, both natural wave attenuation and 
fall down dependence $E_0 (\beta)$ will require very strong $\beta_g$ reduction to the section 
beginning. Such case we have to take larger $\beta_g$ value near section end and lose essentially 
in possible $E_0$ value in the region, where DLW parameters are more attractive.\\
The option of the constant group velocity in BW sections for the system beginning allows much 
higher values of the accelerating gradient and corresponding energy gain for the same length 
of the section. This case a growing along section $E_0$ distribution is realized and it provides 
some additional advantages in particle dynamics.\\
As one will see from examples below, one DLW section in BW mode is not sufficient to overlap low 
$\beta$ region with lower FW efficiency. Two BW sections are required and $\beta_{in} \sim 0.5$ for 
FW part of the system. Direct comparison shows preference for $E_0 = const$ option. It can be 
explained also by plots in Fig. 8. In DLW section with FW mode RF input is in the section beginning, 
where $\beta$ is smaller and, respectively, attenuation is higher and $E_0 (\beta)$ is lower. 
For the case $\beta_g = const$ we spend more RF power in the beginning of the section, where 
DLW parameters are worse.\\        
In the sections design the value of the total attenuation $\tau$ along section is not so important. 
Moreover, some times it can be in contradiction with requirement of the maximal accelerating gradient 
for FW section. In the L-band range the total attenuation $\tau \approx 0.5$ can be obtained either 
with low $\beta_g$, or with lower cell quality factor for lower $\theta$ value, or for large section 
length. As shows previous discussion, see plots in Fig. 7 and Fig. 8, the preferable $\theta$ range for maximal 
energy gain is $\approx (90^o \div 120^0)$ and $Q \geq 10^4$. The $\beta_g$ value is mainly defined 
by requirements of $E_0$ and tuning. If we try to achieve $\tau \approx 0.5$ at the expense of section 
length increasing, it results in reduced RF power in the section end with higher $\beta$ and better 
DLW parameters. It can be more efficient for maximal $E_0$ to stop previous section and start new section 
with higher DLW performance.\\
Also the requirement of higher $E_0$ value in the design is not absolute. In the design of each 
section we scan for optimal $\theta$. It is work for computer. But phase advance, for example $\theta 
=107^o$, even providing slightly higher energy gain as $\theta=90^o$ or $\theta=120^o$, is not 
convenient for realization. Between these two values we chose $\theta=120^o$ due to smaller number of cells 
and lower tuning efforts, (\ref{e11}), if the relative difference in $\delta W$ is not more than $3\%$.     
\subsection{Safety margins}
Before considering examples of DLW accelerating sections, we suggest some safety margins.\\
To calculate the attenuation constant $\alpha$, the quality factor of DLW cells is assumed as $0.95$ from calculated value. 
It takes into account a surface roughness. The cells have a rather simple shape with rotational symmetry and we can 
expect very good quality of surface treatment. Another point is the brazing technique in construction of sections. 
It results normally in additional surface cleaning and in reliable joints between cells. So, the safety margin 
for quality factor for all examples below is accepted as $0.95$.\\
The pulse compression technique is not considered here in details. Referring for the klystron with the output pulse RF power of 
$40 MW$, \cite{keklband}, we assume below that the pulse RF power at the output of pulse compressor is expected as 
$\approx 80 MW$. There will be additional RF losses in transmission and distribution line and and so on. Approximately estimating 
additional RF losses as $\sim 10\%, \approx 8 MW$, we assume that for four accelerating sections will be available from one 
RF source the power $\approx 72 MW$ or $18 MW $ per section. For such input RF power will be considered below all examples 
of accelerating system.\\
Below we consider several examples of DLW accelerating structure design with different $\beta_{g_{ou}}$ values.   
\subsection{The limiting case $\beta_{g_{ou}}=0.04\%$} 
\begin{table}[htb]   
\begin{center}
\centering{Table 2: The first example of the DLW structure design with $\beta_{g_{ou}} =0.4\%$}
\begin{tabular}{|l|c|c|c|c|c|c|c|c|c|c|c|c|c|c|c|c|}
\hline
Section N        & N 1        & N 2        & N 3       & N 4       & N 5       &  N 6      &  N 7      &  N 8        \\
\hline
  Mode           &  BW        & BW         & FW        & FW        &  FW       & FW        &  FW       &  Fw         \\
\hline
$\theta$, deg.   & 90         & 135        & 120       & 120       &  120      & 120       & 120       &  120        \\
\hline
$\beta_{in}$     & 0.08       & 0.28190    & 0.52916   & 0.69935   & 0.79832   & 0.86514   & 0.91432   & 0.92903     \\
\hline
$\beta_{ou}$     & 0.28190    & 0.52916    & 0.69935   & 0.79832   & 0.86514   & 0.90445   & 0.92903   & 0.94522     \\
\hline   
$\phi_s, deg.$   & 30         & 30         & 30        & 30        & 10        & 10        & 10        & 10          \\
\hline   
$\delta W, MeV$  & 4.126      & 14.396     & 23.299    & 27.626    & 36.218    & 37.011    & 37.865    & 38.123      \\
\hline   
$ 2a_{in}, mm$   & 33.132     & 32.920     & 38.082    & 37.684    & 39.902    & 37.860    & 37.796    & 37.730      \\
\hline
$ 2a_{ou}, mm$   & 27.776     & 32.214     & 30.456    & 30.384    & 30.358    & 30.376    & 30.346    & 30.376      \\
\hline
$\beta_{g_{in}},\%$ & 0.4     & 0.4        & 0.834     & 0.822     & 0.0838    & 0.840     & 0.836     & 0.832       \\
\hline
$\beta_{g_{ou}},\%$ & 0.4     & 0.4        & 0.4       & 0.4       & 0.4       & 0.4       & 0.4       & 0.4         \\
\hline
$E_{0_{in}},MV$  & 0.180      & 6.126      & 10.062    & 11.227    & 11.351    & 11.462    & 11.566    & 11.643      \\
\hline
$E_{0_{ou}},MV$  & 4.738      & 9.126      & 10.062    & 11.227    & 11.351    & 11.462    & 11.566    & 11.643      \\
\hline
$\tau, Np$       & 0.76530    & 0.4682     & 0.43973   & 0.4177    & 0.4205    & 0.4205    & 0.4178    & 0.4123      \\
\hline
$Q \cdot 10^{-3}$&$\sim 10.0 $&$\sim 19.5$ &$\sim 14.7$&$\sim 16.8$&$\sim 17.6$&$\sim 18.6$&$\sim 19.0$& $\sim 19.3$ \\
\hline
Length, m        & 2.02       &  2.24      & 2.53      & 2.84      &  3.15     & 3.29      & 3.32      & 3.33        \\       
\hline
N cells          & 76         & 38         & 53        & 49        &  49       & 48        &  47       & 46          \\       
\hline
$Tuning, T_c$    & 18998      &  9499      &  9568     & 8493      &  8296     & 8059      & 7901      & 7734        \\       
\hline
$d_{p_{min}}, mm $& 12.337    &  40.756    & 40.802    & 53.925    & 61.556    & 67.708    & 69.740    & 71.635      \\       
\hline
$t_t, ns$        & 43.981     &  18.326    &  13.672   & 12.603    &  12.603   & 12.346    & 12.088    & 11.831      \\       
\hline
$\tau_f, \mu s$   & 1.69      &  1.87      &  1.54     & 1.65     &  1.79     & 1.84       & 1.87      & 1.87        \\       
\hline
$\frac{E_{smax}}{E_k}$& 1.25  &  1.38      &  0.84     & 0.87      &  0.86     & 0.87      &  0.87     & 0.87        \\       
\hline
\end{tabular}
\label{2t}
\end{center}
\end{table} 
Because the value $\beta_{g_{ou}}=0.04\%$ is the minimal value, which was realized in L-band practice, 
according known references, let us consider it first and start accelerating system with $\beta_{in}=0.08$.
Due to start with very low $\beta$ value and short cell length, even for BW operation, in the first 
section the iris thickness is $t=3.5 mm$. For another sections it is $t=5 mm$.\\
Results are presented in the Table 2,
where $\beta_{in}, 2a_{in}, \beta_{g_{in}}, E_{0_{in}}$ and $\beta_{ou}, 2a_{ou}, \beta_{g_{ou}}, E_{0_{ou}}$ are the input and the 
output values for the particle velocity, the iris aperture, the group velocity and the accelerating gradient of the 
considered DLW section, respectively. Also $\tau$ is the total attenuation, $N$ is the number of cells 
in the section, $d_{p_{min}}$ is the minimal cell length in the section near the input, $t_t, ns$ is the 
traversing time for particles, $\tau_f, \mu s$ is the filling time for the section and $\frac{E_{smax}}{E_k}$ 
is the ration of the maximal electric field at the cell surface to the Kilpatrick threshold 
$E_k \approx 32 \frac{MV}{m}$ for the L-band range.\\
As one can see from the Table 2, the DLW structure with eight section and summarized input RF power of $144 MW$ 
(two klystrons with pulse compressors) overlaps the total energy range of $\sim 218 MeV$. The maximal values 
in FW sections are quite safe, $\frac{E_{smax}}{E_k} \sim 0.8$. For two BW section in the structure 
beginning we obtain enough high $\frac{E_{smax}}{E_k} \sim 1.3$ value, which can require more long 
sections conditioning.\\
This example shows principal possibility to start DLW accelerating structure from $\beta=0.08$, 
but parameters of the first section are difficult for practical realization.\\
Let us consider the first BW section of these example in more details. 
\subsubsection{Parameters of the first section. The second example.}
\begin{figure}[htb] 
\centering
\epsfig{file=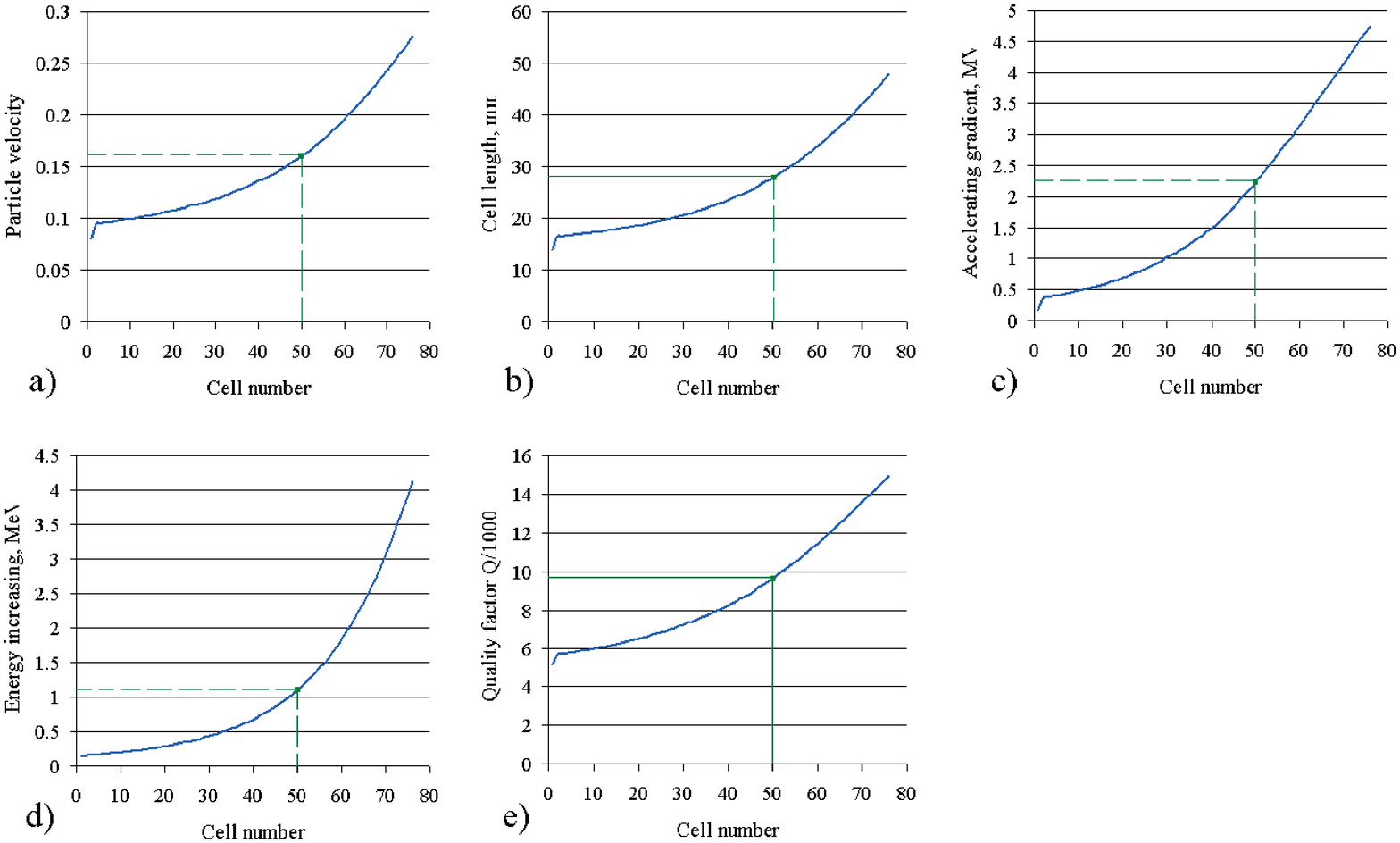, width =150.0mm}
\begin{center}
 Figure 9: For the first BW section, $\beta_g=const =0.4\%$, $\beta_{in}=0.08$, the plots of particle 
velocity $\beta$, (a), the cell length $d_p$, 
(b), the accelerating gradient $E_0$, (c), the achieved energy gain $\Delta W$, (d) and the cell 
quality factor $Q$ (e) in dependence on cell number. 
\end{center}
\label{9f}
\end{figure} 
Some details for particles acceleration in the first BW section with $\beta_{in}=0.08$ are plotted in Fig. 9. Instead of large 
number of cells in the sections, $N=76$, the section beginning with the length $\sim 0.5 m$ and 
with $\sim 50$ cells works for acceleration not effectively. The accelerating gradient is rather low, 
Fig. 9c, (see also Fig. 8a for achievable $E_0$ value) and the energy gain at this section region is 
just $\approx 1 MeV$, Fig. 9d. The velocity of particles rises slow, Fig. 9a, and we have a lot 
of short, Fig. 9b, $d_p \leq 20 mm$ cells. Such cells have relatively low $Q$ values, Fig. 9e, which additionally 
leads to an enlarged wave attenuation.\\   
In the range $0.08 \leq \beta \leq 0.16$ DLW application is not practical. In this very short region 
(required energy gain is just $\sim (1 \div 2) MeV$) is more effective to replace DLW by another accelerating 
structure.\\
If we restrict DLW application for very low $\beta$ and start a structure design from $\beta =0.2$, 
results for the first BW section are much more better. Results for the second example of 
DLW structure design with $\beta_{g_{ou}} =0.4\%$, but starting from $\beta =0.2$, are listed 
in the Table 2a.\\
\begin{table}[htb]   
\begin{center}
\centering{Table 2a: The second example of the DLW structure design with $\beta_{g_{ou}} =0.4\%$}
\begin{tabular}{|l|c|c|c|c|c|c|c|c|c|c|c|c|c|c|c|c|}
\hline
                 & N 1        & N 2        & N 3       & N 4       & N 5       &  N 6      &  N 7      &  N 8        \\
\hline
  Mode           &  BW        & BW         & FW        & FW        &  FW       & FW        &  FW       &  Fw         \\
\hline
$\theta$, deg.   & 120        & 135        & 120       & 120       &  120      & 120       & 120       &  120        \\
\hline
$\beta_{in}$     & 0.2        & 0.45673    & 0.62246   & 0.74982   & 0.82634   & 0.88157   & 0.91432   & 0.93553     \\
\hline
$\beta_{ou}$     & 0.45673    & 0.62246    & 0.74982   & 0.82634   & 0.88157   & 0.91432   & 0.93553   & 0.94973     \\
\hline   
$\phi_s, deg.$   & 30         & 30         & 30        & 30        & 10        & 10        & 10        & 10          \\
\hline  
$\delta W, MeV$  & 10.933     & 16.231     & 24.691    & 27.922    & 36.218    & 37.066    & 38.208    & 38.385      \\
\hline   
$ 2a_{in}, mm$  & 35.130      & 32.236     & 37.490    & 37.382    & 39.902    & 37.744    & 37.816    & 37.728      \\
\hline
$ 2a_{ou}, mm$  & 30.338      & 32.220     & 30.408    & 30.364    & 30.352    & 30.340    & 30.352     & 30.346      \\
\hline
$\beta_{g_{in}},\%$ & 0.4     & 0.4        & 0.802     & 0.802     & 0.0842    & 0.832     & 0.838      & 0.832       \\
\hline
$\beta_{g_{ou}},\%$ & 0.4     & 0.4        & 0.4       & 0.4       & 0.4        & 0.4       & 0.4       & 0.4         \\
\hline
$E_{0_{in}},MV$ & 3.930       & 8.734      & 11.163    & 11.502    & 11.347     & 11.555    & 11.569    & 11.655      \\
\hline
$E_{0_{ou}},MV$ & 8.008       & 9.582      & 11.163    & 11.502    & 11.347     & 11.555    & 11.569    & 11.655      \\
\hline
$\tau, Np$       & 0.4908     & 0.3875     & 0.4088    & 0.4014    & 0.4252     & 0.4164    & 0.4183    & 0.4137      \\
\hline
$Q \cdot 10^{-3}$&$\sim 18.0 $&$\sim 21.9$ &$\sim 16.7$&$\sim 17.8$&$\sim 18.6$&$\sim 18.8$&$\sim 19.3$& $\sim 19.5$ \\
\hline
Length, m        & 2.06       &  2.21      & 2.55      & 2.80      &  3.23     & 3.25      & 3.35      & 3.34        \\       
\hline
N cells         & 41          & 28         & 48        & 46        &  49       & 47        &  47       & 46          \\       
\hline
$Tuning, T_c$  & 10250        &  7000      &  8580     & 8016      &  8253     & 7937      & 7884      & 7742        \\       
\hline
$d_{p_{min}}, mm $ & 30.843   &  66.032    & 35.997    & 57.816    & 63.717    & 67.975    & 70.501    & 72.136      \\       
\hline
$t_t, ns$      & 21.091       &  13.503    &  13.374   & 11.831    &  12.603   & 12.088    & 12.088    & 11.831      \\       
\hline
$\tau_f, \mu s$   & 1.16      &  1.84      &  1.54     & 1.64      &  1.82    & 1.84       & 1.87      & 1.88      \\       
\hline
$\frac{E_{smax}}{E_k}$& 1.33  &  1.37      &  0.87     & 0.88      &  0.86     & 0.87      &  0.87     & 0.87        \\       
\hline
\end{tabular}
\label{2ta}
\end{center}
\end{table} 
Starting DLW acceleration from a higher $\beta_{in} = 0.2$ we ensure much higher accelerating gradient 
in the first DLW BW section, $E_0 \sim (3.93 \div 8.01) \frac{MV}{m}$. It directly leads to energy 
gain increasing, more fast rise of particle velocity, decreasing number of cells in the section, 
decreasing possible tuning efforts $T_c$ and also to decreasing of traversing time. The first cell, 
which has the minimal length in the section, is in length $d_{p_{min}}=30.843 mm$. With the iris 
thickness $t=5 mm$ we have the minimal distance between adjacent irises $\approx 25.5 mm$ and 
it looks sufficient to place in this cell the output RF coupler to direct remaining RF power to external 
RF load. The internal RF load with dissipating layer at the cells surface looks not desirable 
the structure, because will lead to not definite field distribution in cells with distributed RF load.\\
As one can see from the Table 2a, in DLW BW sections remains a high value of field at the 
surface, $\frac{E_{smax}}{E_k} \sim 1.35$.\\
The second example of the structure design also shows the total energy gain $\sim 220 MeV$, which 
can be more, than required. Definitely, we can consider DLW acceleration from $\beta_{in} = 0.16$. 
Because it results in a small change $\sim 1 MeV$ in the energy of input particles, it will 
not change all results strongly. From the plot of accelerating gradient in Fig. 9 one can expect, 
that result for the structure with $\beta_{in} = 0.16$ will be much closer to results with $\beta_{in} = 0.2$ 
than to results with $\beta_{in} = 0.08$.\\
The reserve in the total energy gain $W \sim 220 MeV$ can be spend to relax $\beta_{g_{ou}}$ 
requirement. From (\ref{e9}) one can see, that the ratio $\sqrt{\frac{E_0^2 |\beta_g| }{P_t}}$ 
\begin{equation}
\sqrt{\frac{E_0^2 |\beta_g| }{P_t}} = F(\beta, \lambda, a)
\label{e12}
\end{equation}    
is a some function. In the first approximation, neglecting dependence on aperture radius $a$, from 
each DLW section we can reduce the accelerating gradient by the choice of higher $\beta_g$ value. 
\subsection{The moderate examples $\beta_{g_{ou}}=0.06\%$ and $\beta_{g_{ou}}=0.075\%$}
Supposing that we want to keep approximately the same length of the total structure, as in previous 
structure examples, the third option with $\beta_{g_{ou}} =0.6\%$ has been considered. The results 
of this third examples are listed in the Table 3.\\ 
\begin{table}[htb]   
\begin{center}
\centering{Table 3: The DLW structure design with $\beta_{g_{ou}} =0.6\%$}
\begin{tabular}{|l|c|c|c|c|c|c|c|c|c|c|c|c|c|c|c|c|}
\hline
                 & N 1        & N 2        & N 3       & N 4       & N 5       &  N 6      &  N 7      &  N 8        \\
\hline
  Mode           &  BW        & BW         & FW        & FW        &  FW       & FW        &  FW       &  Fw         \\
\hline
$\theta$, deg.   & 120        & 135        & 120       & 120       &  120      & 120       & 120       &  120        \\
\hline
$\beta_{in}$     & 0.2        & 0.40072    & 0.56196   & 0.70265   & 0.79098   & 0.85533   & 0.89447   & 0.91990     \\
\hline
$\beta_{ou}$     & 0.40072    & 0.56196    & 0.70265   & 0.79098   & 0.85533   & 0.89447   & 0.91990   & 0.93741     \\
\hline   
$\phi_s, deg.$   & 30         & 30         & 30        & 30        & 10        & 10        & 10        & 10          \\
\hline  
$\delta W, MeV$  & 7.485      & 12.414     & 20.756    & 24.197    & 31.253    & 32.369    & 33.130    & 33.991      \\
\hline   
$ 2a_{in}, mm$  & 36.228      & 36.568     & 40.574    & 40.264    & 40.470    & 40.354    & 40.308    & 40.344      \\
\hline
$ 2a_{ou}, mm$  & 34.234      & 36.336     & 34.404    & 34.262    & 34.210    & 34.216    & 34.214    & 34.200      \\
\hline
$\beta_{g_{in}},\%$ & 0.6     & 0.6        & 1.018     & 1.014     & 1.038     & 1.032     & 1.030     & 1.034       \\
\hline
$\beta_{g_{ou}},\%$ & 0.6     & 0.6        & 0.6       & 0.6       & 0.6       & 0.6       & 0.6       & 0.6         \\
\hline
$E_{0_{in}},MV$ & 2.494       & 6.092      & 9.396     & 9.891     & 9.976     & 10.140    & 10.227    & 10.253      \\
\hline
$E_{0_{ou}},MV$ & 5.676       & 7.211      & 9.396     & 9.891     & 9.976     & 10.140    & 10.227    & 10.253      \\
\hline
$\tau, Np$       & 0.3375     & 0.2773     & 0.3207    & 0.3085    &  0.3178   & 0.3103    & 0.3072    & 0.3089      \\
\hline
$Q \cdot 10^{-3}$&$\sim 17.5 $&$\sim 20.7$ &$\sim 15.9$&$\sim 17.3$&$\sim 18.2$&$\sim 18.8$&$\sim 19.1$& $\sim 19.3$ \\
\hline
Length, m        & 2.05       &  2.24      & 2.55      & 2.83      &  3.18     & 3.24      & 3.29      & 3.37        \\       
\hline
N cells         & 45          & 32         & 52        & 49        &  50       & 48        &  47       & 47          \\       
\hline
$Tuning, T_c$  & 7500         &  5330      &  6980     & 6335      &  6305     & 6024      & 5885      & 5865        \\       
\hline
$d_{p_{min}}, mm $ & 30.843   &  57.934    & 47.996    & 54.179    & 60.990    & 65.952    & 68.970    & 70.931      \\       
\hline
$t_t, ns$        & 23.148     &  15.432    & 12.346    & 12.603    &  12.860   & 12.346    & 12.088    & 12.088      \\       
\hline
$\tau_f, \mu s$   & 1.16      &  1.25      &  1.15     & 1.22      &  1.40     & 1.36      & 1.38      & 1.40        \\       
\hline
$\frac{E_{smax}}{E_k}$& 1.11  &  1.17      &  0.81     & 0.81      &  0.81     & 0.80      &  0.81     & 0.81        \\       
\hline
\end{tabular}
\label{3t}
\end{center}
\end{table} 
This example looks as mostly balanced for muons acceleration, because the main works for energy gain is 
shifted to DLW FW operation and the load between FW sections, which can be roughly reflected by 
attenuation $\tau$ is distributed more or less uniform, following to the DLW performance change 
with $\beta$, see plots in Fig. 8 and Fig. 7. The phase advance for section is chosen near the 
optimal for the maximal energy gain, but keeping in mind the minimal cell length in the section and also 
taking into account the reduction of cell number with higher $\theta$. For last sections in FW part 
the optimal $\theta$ value is closer to $\theta =90^o$. But the dependence is very smooth, see Fig. 7a, 
and loosing near $2\%$ in the energy gain along FW part, with $\theta=120^o$ we decrease the number 
of cells and related tuning efforts at least at $\sim (18 \div 20)\%$.\\
Due to general decreasing of accelerating gradient in the BW part, the maximal field value 
$\frac{E_{smax}}{E_k} \leq 1.2$ comes to quite acceptable level.\\  
Further increasing of $\beta_{g_{ou}}$ value with is possible at the expense of increasing of the length of DLW sections.
 In the Table 4 are listed the results of structure for $\beta_{g_{ou}} =0.75\%$. In this 
forth example we keep approximately the same energy gain for each DLW section with respect to the 
optimal third example, see Table 3.\\
\begin{table}[htb]   
\begin{center}
\centering{Table 4: The DLW structure design with $\beta_{g_{ou}} =0.75\%$ at the expense of increased length}
\begin{tabular}{|l|c|c|c|c|c|c|c|c|c|c|c|c|c|c|c|c|}
\hline
                 & N 1        & N 2        & N 3       & N 4       & N 5       &  N 6      &  N 7      &  N 8        \\
\hline
  Mode           &  BW        & BW         & FW        & FW        &  FW       & FW        &  FW       &  Fw         \\
\hline
$\theta$, deg.   & 120        & 135        & 120       & 120       &  120      & 120       & 120       &  120        \\
\hline
$\beta_{in}$     & 0.2        & 0.39989    & 0.56033   & 0.70167   & 0.79204   & 0.85581   & 0.89526   & 0.92087     \\
\hline
$\beta_{ou}$     & 0.39989    & 0.56033    & 0.70167   & 0.79204   & 0.85581   & 0.89526   & 0.92087   & 0.93813     \\
\hline   
$\phi_s, deg.$   & 30         & 30         & 30        & 30        & 10        & 10        & 10        & 10          \\
\hline  
$\delta W, MeV$  & 7.440      & 12.288     & 20.728    & 24.785    & 31.179    & 32.886    & 32.260    & 34.114      \\
\hline   
$ 2a_{in}, mm$  & 39.318      & 39.234     & 42.882    & 42.644    & 42.634    & 42.620    & 42.478    & 42.526      \\
\hline
$ 2a_{ou}, mm$  & 36.642      & 39.878     & 36.832    & 36.658    & 36.602    & 36.596    & 36.580    & 36.588      \\
\hline
$\beta_{g_{in}},\%$ & 0.75    & 0.75       & 1.206     & 1.214     & 1.224     & 1.238     & 1.218     & 1.224       \\
\hline
$\beta_{g_{ou}},\%$ & 0.75    & 0.75       & 0.75      & 0.75      & 0.75      & 0.75      & 0.75      & 0.75         \\
\hline
$E_{0_{in}},MV$ & 1.888       & 5.132      & 8.427     & 8.843     & 9.039     & 9.149     & 9.270     & 9.292      \\
\hline
$E_{0_{ou}},MV$ & 4.830       & 6.231      & 8.427     & 8.843     & 9.039     & 9.149     & 9.270     & 9.292      \\
\hline
$\tau, Np$       & 0.3311     & 0.2563     & 0.2970    & 0.2897    &  0.2889   & 0.2869    & 0.2803    & 0.2813      \\
\hline
$Q \cdot 10^{-3}$&$\sim 17.5 $&$\sim 20.7$ &$\sim 15.9$&$\sim 17.3$&$\sim 18.2$&$\sim 18.8$&$\sim 19.1$& $\sim 19.3$ \\
\hline
Length, m        & 2.48       &  2.58      & 2.84      & 3.24      &  3.50     & 3.65      & 3.64      & 3.73        \\       
\hline
N cells         & 55          & 37         & 58        & 56        &  55       & 54        &  52       & 52          \\       
\hline
$Tuning, T_c$  & 7332         &  4933      &  6453     & 5954      &  5738     & 5577      & 5375      & 5348        \\       
\hline
$d_{p_{min}}, mm $ & 30.843   &  57.814    & 43.205    & 54.104    & 61.072    & 65.989    & 69.037    & 71.006      \\       
\hline
$t_t, ns$      & 28.292       &  17.843    & 14.918    & 14.403    &  14.146   & 13.889    & 13.374    & 13.374      \\       
\hline
$\tau_f, \mu s$   & 1.08      &  1.16      &  1.06     & 1.15      &  1.22     & 1.26      & 1.26      & 1.28        \\       
\hline
$\frac{E_{smax}}{E_k}$& 1.01  &  1.07      &  0.74     & 0.75      &  0.75     & 0.75      &  0.75     & 0.75        \\       
\hline
\end{tabular}
\label{4t}
\end{center}
\end{table} 
For the forth example, as one can see from the Table 4, with $\beta_g$ increasing in the first BW 
sections we also can get quite safe values for maximal electric field at the surface.\\ 
Definitely, an additional example, fifth, can be combined by taking the structure beginning with BW sections
from the forth example, see Table 4, and continuing with FW sections from the third example, see 
Table 3. Such case we relax the disadvantage of the third example in the high surface electric 
field in BS sections and take advantage of the shorter structure part with FW sections.\\   
The previous examples were designed supposing two klystrons with pulse compressors and eight 
accelerating sections. At the expense of RF power we can consider more options, supposing 
DLW sections with larger group velocity.
\subsubsection{DLW system with three RF sources.}
Let us consider options of DLW sections with a twice higher RF pulse power of $36 MW$. Such 
design of DLW section can require design of a symmetrical RF input coupler to divide enough high 
input pulse power between two shoulders of the coupler.\\
\begin{table}[htb]   
\begin{center}
\centering{Table 5: The DLW structure design for three RF source with the total RF pulse power 
 $216 MW$}
\begin{tabular}{|l|c|c|c|c|c|c|c|c|c|c|c|c|c|c|c|c|}
\hline
                 & N 1        & N 2        & N 3              & N 4              & N 5              &  N 6             \\
\hline
  Mode           &  BW        & BW         & FW               & FW               &  FW              & FW               \\
\hline
$\theta$, deg.   & 120        & 150        & 90 (120)         & 90 (120)         & 90 (120)         & 90 (120)         \\
\hline
$\beta_{in}$     & 0.2        & 0.4224    & 0.6199           & 0.7742(0.7726) & 0.8613(0.8615) & 0.9093(0.9080) \\
\hline
$\beta_{ou}$     & 0.4224     & 0.6199     & 0.7742(0.7726)   & 0.8613(0.8615) & 0.9093(0.9080) & 0.9370(0.9358) \\
\hline   
$\phi_s, deg.$   & 30         & 30         & 30               & 10               & 10               & 10               \\
\hline  
$\delta W, MeV$  & 9.976      & 16.834     & 32.272(31.780)   & 41.505(41.640)   & 45.929(45.397)   & 48.854(47.541)   \\
\hline   
$ 2a_{in}, mm$  & 42.188      & 46.722     & 42.658(43.236)   & 42.090(42.998)   & 42.182(43.060)   & 42.284(43.112)   \\
\hline
$ 2a_{ou}, mm$  & 38.684      & 45.842     & 35.940(37.438)   & 35.802(37.336)   & 35.750(37.302)   & 35.724(37.286)   \\
\hline
$\beta_{g_{in}},\%$ & 0.9     & 0.9        & 1.374 (1.262)    & 1.344 (1.258)    & 1.364 (1.270)    & 1.380 (1.278)    \\
\hline
$\beta_{g_{ou}},\%$ & 0.9     & 0.9        & 0.8 (0.8)        & 0.8 (0.8)        & 0.8 (0.8)        & 0.8 (0.8)        \\
\hline
$E_{0_{in}},MV$ & 2.072       & 6.717      & 12.366(12.091)   & 13.051(12.600)   & 13.173(12.736)   & 13.207(12.801)   \\
\hline
$E_{0_{ou}},MV$ & 6.374       & 8.486      & 12.366(12.091)   & 13.051(12.600)   & 13.173(12.736)   & 13.207(12.801)   \\
\hline
$\tau, Np$       & 0.2660     & 0.2085     & 0.3231(0.2772)   & 0.3011(0.2673)   &  0.3060(0.2699)  & 0.3108(0.2718)   \\
\hline
Length, m        & 2.51       &  2.60      & 3.01(3.03)       & 3.23(3.35)       &  3.55(3.62)      & 3.74 (3.77)      \\       
\hline
N cells         & 52          & 36         & 74(56)           & 68(53)           &  69 (53)         & 70 (53)          \\       
\hline
$Tuning, T_c$  & 5777         &  3999      &  7375(5832)      & 6604(5324)       &  6555(5233)      & 6571(5189)       \\       
\hline
$d_{p_{min}}, mm $ & 30.843   &  59.692    & 35.847(47.796)   & 44.681(59.575)   & 49.808(66.279)   & 52.283(70.012)   \\  
\hline
$t_t, ns$      & 26.749       &  16.204    & 14.275(14.403)   & 13.117(13.632)   &  13.310(13.632)  & 13.503(13.623)   \\       
\hline
$\tau_f, \mu s$   & 0.90      &  0.97      &  1.01(1.06)      & 1.05(1.13)       &  1.12 (1.19)     & 1.17 (1.23)      \\       
\hline
$\frac{E_{smax}}{E_k}$& 1.33  &  1.33      &  0.87(1.04)      & 0.90(1.05)       &  0.90 (1.05)     & 0.90 (1.05)      \\       
\hline
\end{tabular}
\label{5t}
\end{center}
\end{table} 
This example will have just six accelerating sections and increased, as in previous examples, the 
group velocity value. Even with increased RF power, a sufficient simultaneous rise both in accelerating 
gradient $E_0$ and in $\beta_g$ is not possible, see (\ref{e12}). One FW section is not sufficient 
to put particles in $\beta$ region with preferable FW operation. And two BW sections is a forced 
solution. We design these two BW sections with the constant group velocity $|\beta_g| =0.9\%$. 
The following FW part is considered in two options - with operating phase advance $\theta=90^o$, to have 
the maximal energy gain, and with $\theta=120^o$, to have smallest number of cells and decreased 
tuning efforts. Results are listed in the Table 5 with results for $\theta=120^o$ in brackets.\\  
One can clearly compare from the Table 5 two options for FW part. The difference in energy gain 
between options $\theta=90^o$ and $\theta=120^o$ is compensated just by three additional cells and results 
in the structure length increasing at $0.24 m$ at the background of total length $\sim 18.64 m$. 
The maximal electric field at the surface is higher for $\theta=120^o$ but is in the safe limits. 
But total number of cells and corresponding tuning for $\theta=90^o$ are evidently higher. 
\subsection{Results consideration and discussion}
For more clear comparison the main parameters of examples, considered above, are summarized in the 
Table 6.\\
\begin{table}[htb]   
\begin{center}
\centering{Table 6: The summary for examples of accelerating structures with DLW sections.}
\begin{tabular}{|l|c|c|c|c|c|c|c|c|c|c|c|c|c|c|c|c|}
\hline
Example $N$                    & 1          &  2          & 3       & 4       &   5       &   6     \\
\hline
$P_{tot}=\sum P_t, MW$         & 144        &  144        & 144     & 144     & 144       & 216     \\
\hline
$\beta_{g_{ou}},\%$            & 0.4        & 0.4         & 0.6     & 0.75    & 0.75-0.6  & 0.9-0.8 \\
\hline
$\beta_{in}$                   & 0.08       & 0.2         & 0.2     & 0.2     & 0.2       & 0.2     \\
\hline
$\beta_{ou}$                   & 0.94522    & 0.94973     & 0.93741 & 0.93813 & 0.93813   & 0.93702(0.93579)\\
\hline   
$\Delta W = \sum \delta W, MeV$& 218.66     & 229.65      & 195.60  & 195.7   & 195.65    & 195.37 (193.17) \\
\hline  
$Length, L=\sum l$, m          & 22.72      & 22.79       & 22.75   & 25.66   & 23.52     & 18.64 (18.88)   \\       
\hline
$N cells, N_t=\sum N$          & 362        & 352         & 370     & 420     & 385       & 369 (303)       \\       
\hline
$N_{sectios}$                  &    8       &    8        &   8     &   8     &   8       &    6            \\
\hline
$Tuning, T_c^t =\sum T_c$      & 78549      &  65662      & 50224   & 46710   & 45804     & 36881 (31354)   \\       
\hline
$T_f=\sum t_f, ns$             & 125.62     &  108.41     & 112.92  & 130.24  & 120.475   & 98.158 (99.243) \\       
\hline
\end{tabular}
\label{6t}
\end{center}
\end{table}
As it should be for the L-band operating frequency, see Chapter 2.2, tolerable options can be 
obtained at the expense of high input RF power and not so high group velocity. But, starting from 
example $N=3$ we consider $\beta_g$ for accelerating sections, which is \textbf{not less} than the 
maximal $\beta_g$ value in already realized references, \cite{keklband}. As one can see from the 
summarizing Table 6, required number of DLW sections is not so big. At least, with $8$ sections 
the goal $W \sim 195 MeV$ can be achieved assuming $\beta_g \geq 0.6$. it is not so small group 
velocity, even for the S-band range. It may me mentioned here, that in the classical S-band constant gradient accelerating 
sections, with 10 foot length, in the Stanford two-mile accelerator the group velocity changes 
as $2.04\% \geq \beta_g \geq 0.65\%$.\\
Also consideration shows variety of different possibilities. In the practical linac design all parameters 
should be balanced. The results, presented in the Table 6 show some limitations and some trends 
of results change in dependence on input data.\\
Results of study show some practical benefits for FW DLW sections realization with $\theta=120^o$. 
With a quite small loss $\sim 3\%$ in the energy gain one can obtain quite visible reduction 
of efforts in the tuning of accelerating sections. Another attractive feature is larger cell length, 
which, together with the reduced thickness of the iris, provides comfortable spacing between 
adjacent irises for low particles velocity.  
\section{Some remarks for beam dynamics.}
The particles dynamics is not considered in this work, just very approximately a longitudinal 
bunch length and longitudinal oscillations are taken in consideration by the synchronous phase 
$\phi_s \sim (30^o \div 10^o)$, assuming phase length reduction with energy increasing.\\
As show above results of RF properties estimations, acceleration of muons with a relative high frequency 
DLW structure is possible (or practically reasonable) from low energy $W_m \approx 107.84 MeV$. 
Below this energy enough narrow range $\approx 3 MeV$ should be covered with another structures, with 
lower operating frequency. The matching in longitudinal motion will be required. Such technique is 
known in high energy proton linacs.\\
More interesting is the problem of particles focusing and one possibility for BW DLW sections should 
be mentioned. During our study of RF parameters this possibility was indicated for us by Dr. R. Jameson.\\
Now there are well known two methods of focusing with RF fields - Radio Frequency Quadrupole (RFQ) 
and Alternating Phase Focusing (APF). There is the third idea for focusing with RF field, especially favorable for structures 
operating at higher spatial harmonic. In our case of BW DLW the synchronous harmonic $n=-1$, 
see (\ref{e1}), is not fundamental. In such DLW structure all time exists harmonic $n=0$ with larger 
amplitude and higher phase velocity. As it was shown in \cite{tkalich}, such fast 
harmonic can provide the focusing effect. This effect was investigated theoretically for focusing in 
proton accelerators, see, for example \cite{baev} with appropriate references, and some later papers.
In practical realization for electron linac, \cite{aizaz}, the pulse current up to $2 A$ was observed 
in the backward  wave DLW section, $\theta=120^o$, without external focusing.\\
When the backward wave operating mode is specially selected to use the focusing effect of the fast 
harmonic, the price for it is reduced RF efficiency of accelerating structure (similar to APF and RFQ).
For muons acceleration this backward mode is suggested to extend the same structure - DLW - to middle 
and low range of particle velocity. And we have already the fast harmonic in backward sections as the sequence 
of another reasons. The beam intensity for muons is negligible and space charge effects are practically 
zero. Such case we have more freedom in the radial focusing, as compared to \cite{baev}, \cite{aizaz},
 and we should provide just conditions for a stable transverse motion and tolerable size of beam 
envelope.\\
It looks interesting at least to estimate such possibility for muons acceleration. If the focusing 
effect of the fast harmonic will be sufficient, the particles focusing in the beginning of the 
structure will be free of additional charge, just as the sequence of DLW BW application. 
\section{Summary}
The well known accelerating structure - the Disk Loaded Wavequide - is considered in details for 
applications in the L-band frequency range and for acceleration of particles with intermediate 
mass in the wide range of particles velocity. For the L-band range the dimensions of cells can be 
optimized in the iris thickness to have higher accelerating gradient, lower attenuation and larger 
space between irises. By choosing operating phase advance $\theta=120^o$, one can propose DLW 
accelerating system, which overlap simultaneously both high particle energy range and moderate energy range too. 
With application of the backward wave operating mode we can extend DLW structure to low range of particles 
velocity. \\
Combining BW operating mode in the system beginning and FW operating mode for medium and high particles 
energy, we can suggest DLW accelerating system, which overlap the muons energy range from $\beta=0.2$ 
to $\beta \sim 0.93$  with $8$ or $6$ accelerating sections, just using two or three RF sources with 
RF pulse compression.\\
As compared to another possible solutions for low and medium energy range, such accelerating 
system has the preferences of uniformity, simplicity and cost efficiency.   
\section{Acknowledgments}
The author thanks Dr. R. Jameson for his comments about focusing in a backward wave structure, Dr. M. Yoshida, KEK, 
for his critical comments about DLW application at low particle velocity, Dr. S. Polozov, MEPhI, for 
his interest and discussion in the beam dynamic issue. I also thank all another my Colleagues for interest, 
stimulation and comments in this study.


\begin{thebibliography}{99}
\addcontentsline{toc}{section}{\refname}
\bibitem{wanglband} J. W. Wang et. al., Studies of room temperature structures for the ILC positron source.
Proc. 2005 PAC, p. 2827, 2005
\bibitem{keklband} S. Matsumoto et. al., L-band accelerator system in injector linac for SuperKEKB.
Proc. IPAC 2010, p. 3708, 2010
\bibitem{compend} J. Clendenin et. al., Compendium of scientific linacs. CERN-PS 96-32, Proc. 1996 Linac, 1996
\bibitem{lapost} G. Loew, R. Neal,  Accelerating Structures, G. Dome, Review and Survey of Accelerating 
Structures, in Linear Acceletors, ed. P. Lapostolle, E. Septier, Amsterdam, 1970\\
 G. Loew, R. Neal et. al, The Stanford two-mile accelerator, W. Benjamin Inc., New York, Amsterdam, 1968
\bibitem{aizaz} M.I. Aizatsky et. al., High current electron linac for investigations in new methods 
of acceleration. Fizika plazmy, Nauka, v. 20, n. 7,8, p. 671, 1994 (in Russian).\\ 
A.N. Dovbnya et. al., Beam parameters of an S-band electron linac with beam energy of 30..100 MeV.
Problems of Atomic Science and Technology, Series: Nuclear Physics Investigations. Kharkiv, Ukraine, 
(v. 46), n. 2, p.11-13, 2006. 
\bibitem{library} V. Paramonov, The data library for accelerating structures development. ...
Proc. 1996 Linac, p. 493, 1996
\bibitem{gella} I. Gonin et. al., 2D codes set for RF cavities design. Proc. 1990 EPAC, p. 1026, 1990.
\bibitem{lanlcode} F. Krawczyk et. al., The Los Alamos accelerator code group. Proc. 1995 PAC, p. 2306, 1996.
\bibitem{tuning} T.Khabiboulline, V.Puntus et. al., A new tuning method for traveling wave structures. 
Proc. 1995 PAC, p. 1666
\bibitem{pulse} M.Yoshida, T. Shintake. Efficiency and gain enhancement of RF-pulse compressor 
for C-band RF system. Proc. 1998 Linac Conf., p. 935.
\bibitem{muon} S. Artikova, F. Naito, M. Yoshida. The accelerator design of muon g-2 experiment at 
J-PARC. Proc. 2013 IPAC, p.1334\\
Conceptual design report for the measurement of the muon anomalous magnetic moment g-2 and electric 
dipole moment at J-PARC, December 13, 2011.
\bibitem{tkalich} V. Tkalich, Zh. Eksp. Teor. Fiz., v. 32, p. 625, 1957. Sov. Phis. JETP, v. 5, 1957 
\bibitem{baev} V. Baev, S, Minaev. Efficiency of ion focusing by the field of a traveling wave in 
a linear accelerator. Zh. Tekh. Fiz., v. 51, p. 2310, 1981. Sov. Phys. Tech. Phys., v. 26(11), 1981  
\end{thebibliography}
\end{document}